\documentclass[preprint,11pt]{elsarticle}
\usepackage{amssymb,amsmath,amsthm}
\usepackage{algorithm}
\usepackage{algpseudocode}
\usepackage{multirow}
\usepackage{float}
\usepackage{todonotes}
\usepackage{xcolor}
\usepackage{natbib,hyperref}
\usepackage{longtable}

\def\bs{\boldsymbol}

\newcounter{algsubstate}


\newtheorem{definition}{Definition}%

\journal{Journal of Signal Processing Systems}

\begin{document}

\begin{frontmatter}

\title{Efficient Image Denoising by Low-Rank Singular Vector Approximations of Geodesics' Gramian Matrix}

\author[1]{Kelum Gajamannage\corref{cor1}}\ead{kelum.gajamannage@tamucc.edu}
\author[2]{Yonggi Park}\ead{yonggi.park@tamucc.edu}
\author[2]{S. M. Mallikarjunaiah}\ead{M.Muddamallappa@tamucc.edu}
\author[3]{Sunil Mathur}\ead{kmathur111@yahoo.com}
\cortext[cor1]{Corresponding author}

\address[1]{Department of Mathematics and Applied Mathematical Sciences, University of Rhode Island, 45 Upper College Rd., Kingston, RI-02881, USA}
\address[2]{Department of Mathematics and Statistics, Texas A\&M University - Corpus Christi, 6300 Ocean Dr., Corpus Christi, TX-78412, USA}

\begin{abstract}
With the advent of sophisticated cameras, the urge to capture high-quality images has grown enormous. However, the noise contamination of the images results in substandard expectations among the people; thus, image denoising is an essential pre-processing step. While the algebraic image processing frameworks are sometimes inefficient for this denoising task as they may require processing of matrices of order equivalent to some power of the order of the original image, the neural network image processing frameworks are sometimes not robust as they require a lot of similar training samples. Thus, here we present a manifold-based noise filtering method that mainly exploits a few prominent singular vectors of the geodesics' Gramian matrix. Especially, the framework partitions an image, say that of size $n \times n$, into $n^2$ overlapping patches of known size such that one patch is centered at each pixel. Then, the prominent singular vectors, of the Gramian matrix of size $n^2 \times n^2$ of the geodesic distances computed over the patch space, are utilized to denoise the image. Here, the prominent singular vectors are revealed by efficient, but diverse, approximation techniques, rather than explicitly computing them using frameworks like Singular Value Decomposition (SVD) which encounters $\mathcal{O}(n^6)$ operations. Finally, we compare both computational time and the noise filtration performance of the proposed denoising algorithm with and without singular vector approximation techniques.
\end{abstract}

\begin{keyword}
Efficient algorithm \sep noise filtration \sep singular vector approximation  \sep geodesics' Gramian
\end{keyword}

\end{frontmatter}

%%%%%%%%%%%%%%%%%%%%%%%%%
%%%%%%%   Introduction    %%%%%%%%%
%%%%%%%%%%%%%%%%%%%%%%%%%
\section{Introduction}
The surge of high-caliber imaging devices in fields such as medical diagnosis, astronomy, and the film industry produces images with more pixels per inch which are in top-notch condition  \cite{GajamannageGGDwtNoise}. However, because of unforeseen instrumental and environmental conditions, the images generated by them might be contaminated with noise \cite{Fan2019}. Thus, filtration of noise to restore the natural image quality that includes all the image features including edges and texture is required as a prepossessing step. Especially, such noise filtration is essential if the images are used for any subsequent processes such as remote sensing or object tracking. In any quality image, the high-frequency components are the noise, edges, and texture, then discerning them to filter out noise is an arduous task \cite{wang2020}. Such a disadvantage periodically brings about the loss of key characters of the retrieved image. Therefore restoration of the original quality of the recovered images without losing essential features is a crucial property that denoising methods should possess. There are mainly two types of image denoising: the patch-based method and the pixel-based method. The advantages of the former method are twofold: first, smoothening of the predominantly ``flat region'' due to the overlap among the patches; second, the capability of preserving the details of the fine image and sharp edges \cite{alkinani2017patch}.

Notably, there are mainly three image denoising methods based on the deep learning neural network approach: DRUNET- Dilated Residual U-Net  \cite{Devalla2018}; residual learning by deep convolution neural network method \cite{Zhang2017}; fast and flexible convolution neural network based denoising \cite{Zhang2018}. These methods mainly learn a \textit{mapping function} on a training data set containing the clean image pairs by optimizing the total loss function \cite{Fan2019}. The aforementioned methods have received a lot of attention recently mainly due to their tremendous success in image denoising and computer vision  \cite{Fan2019}. However, these methods have several major drawbacks such as complications in training when the noise contamination is very high, the issues with the vanishing gradient when the network is considerably ``deep'', and the expensive computations needed for the repeated training process  \cite{gajamannage2022reconstruction}. 

The proposed framework uses the singular vectors of the Gramian matrix of geodesic distances between patches for denoising which we denote as Geodesic Gramian Denoising (GGD). In particular, the $n\times n$ noisy image is partitioned into moderately overlying patches that are square-shaped and with known length $\rho$. The discretization of the image is done in such a way that each patch is centered at a unique pixel. Such partition results in $n^2$ number of patches as points in $\rho^2$-dimensional space with an underlying low-dimensional manifold \cite{Gajamannage2021, SGE}. Then, we compute the geodesic distance matrix $\mathcal{D}$ with size $n^2\times n^2$ which is later converted into the corresponding Gramian matrix where the singular vectors of the Gramian matrix are utilized to produce the noise-filtered version of the given image. Use of Singular Value Decomposition (SVD) for explicit computation of singular vectors of GGD, denoted as GGD-SVD, encounters high computational complexity of $\mathcal{O}(n^6)$ when denoising an image of size $n\times n$ since it requires the computation of SVD of a Gramian matrix of size $n^2\times n^2$. Numerical implementation of SVD is carried out using the linear algebra package LAPACK routine available in Ref.~\cite{lapack99} that includes two processes: first, the given  $m \times n$ matrix is reduced to a bidiagonal form using  \textit{Householder transformations} with $\mathcal{O}(mn \times \min(m, n))$ operations, and then use the QR-algorithm in Ref.~\cite{Francis1961, Francis1962}, to compute the SVD of the bidiagonal matrix \cite{Golub1970}. Thus, instead of the explicit computation of singular vectors of the Gramina matrix; here, we approximate them using four iterative schemes.

The \textit{Monte Carlo Low-Rank Approximation}  (MCLA) \cite{friedland2006fast}, iteratively approximates a $L$-rank surrogate for the conventional SVD of a data matrix $A$ of size $m\times n$. The computational complexity of MCLA is $O(kmn)$, since  each iteration is associated with the reading of $\mathcal{O}(L)$ of columns or rows of $A$. One of the most prominent distinctions of MCLA with other well-known algorithms is that at each iteration the low-rank approximation is guaranteed to be better compared to any previous iteration. The  \textit{Augmented Lanczos Bidiagonalization}  (ALB) \cite{baglama2005augmented}, is another algorithm for the low-rank approximation, which primarily computes some of the largest or smallest singular triplets of a matrix. The procedure in ALB primarily computes a sequence of projections on low-dimensional Krylov subspaces, similar to the procedure in the Lanczos bidiagonalization method. The \textit{Preconditioned Iterative Multimethod Eigensolver} (PIME) \cite{wu2017primme_svds}, provides a preconditioned, two-stage approach to efficiently and meticulously approximate a small number of extreme singular triplets. The PIME solver,  which is developed on the platform of a well-known \textit{eigensolver} PRIMME, for the rages of singular values of the matrix $A$. The \textit{Randomized Singular Value Decomposition} (RSVD) \cite{halko2011finding}, is a powerful tool that computes partial matrix decompositions and then random samples of those decompositions to identify a subspace that captures most features of a matrix. Now, instead of the explicit SVD computation of the Gramian matrix, we reveal singular vectors using the four approximation frameworks, MCLA, ALB, PIME, and RSVD.

The organization of the current paper is as follows: in Sec.~\ref{sec:method}, details about GGD-SVD, MCLA, PRIME, and RSVD  are provided; then, the comparison of the denoising performance of GGD-SVD and those four hybrid GGD frameworks, GGD-MCLA, GGD-ALB, GGD-PIME, and GGD-RSVD, with respect to their parameters using the three reconstruction metrics PSNR, SSIM, and RE is done in  Sec.~\ref{sec:analysis}. Finally, a summary along with conclusions are provided in Sec.~\ref{sec:conclusion}. Then, Tables~\ref{tab:nome} and Table~\ref{tab:abrv} provides the nomenclature and abbreviations, respectively.

%%%%%%%%%%%%%%%%%%%%%%%%%
\begin{table}[htp]
\caption {Nomenclature} \label{tab:nome}
\begin{tabular}{p{1.2cm}|p{11cm}}
Notation &  Description \\

\hline %scaler :
$\sigma_i$ & $i$-th Singular value\\
$\lambda_l$ & $l$-th eigenvalue \\
$\epsilon$ & Machine epsilon\\
$\eta$ & Tolerance \\
$\alpha$ & Relative Norm of Noise\\
$\zeta$ & Relative noise \\

%GGD :
$d(k, k')$ & Distance between patches $\bs{u}(\bs{x}_{k})$ and $\bs{u}(\bs{x}_{k'})$\\
$L$ & Singular triplets' threshold such that $l=1,\dots,L$\\
$L'$ & Singular triplets' threshold of ALB such that $\{\hat{\sigma}_i, \hat{\bs{u}}_i, \hat{\bs{v}}_i \}, i=1,...,L'$ \\
$i$,$j$, $k$ & Indices for matrices \\
$l$ & Index for singular triplets. \\
$(t)$ & Index for iterations of an algorithm. \\
$k$ & $ij$-th pixel's single reference index where $k=n(i-1)+j$. \\
$m$ & Number of rows of a matrix or length of an image \\
$n$ & Number of columns of a matrix or width of an image \\
$h$ & Number of rows or columns of a matrix \\
$\delta$ & Nearest neighbor parameter \\
$\rho$ &  Patch size \\
%$\Delta$ & Reconstruction error \\

\hline %matrix :
$U$ & $= [\bs{u}_1, ..., \bs{u}_n] \in \mathcal{R}^{m \times n}$\\
$V$ & $= [\bs{v}_1, ..., \bs{v}_n] \in \mathcal{R}^{m \times n}$\\
$\tilde{V}$ & $=[\bs{\nu}_1\vert \dots \vert \bs{\nu}_l \vert \dots \vert \bs{\nu}_{n^2}]^T$\\
$\Lambda$ & $= Diag(\lambda_1, \dots, \lambda_l, \dots, \lambda_{n^2})$\\
$\Sigma$ & $= Diag({\sigma_1, ..., \sigma_n})$\\

$\tilde{U}_{\ell+1}$ & $= [\bs{p}_1, ..., \bs{p}_{\ell+1}] \in \mathcal{R}^{m \times (\ell+1)}$, Matrix in ALB\\
$U_{\ell+1}$ & $= [\bs{q}_1, ..., \bs{q}_{\ell+1}] \in \mathcal{R}^{n \times (\ell+1)}$, Matrix in ALB\\
$\hat{\tilde{U}}_{k+1}$ & Predicted of $\tilde{U}_{\ell+1}$ for (k+1)-th singular triplets\\
$\hat{U}_{k+1}$ & Predicted of $U_{\ell+1}$ for (k+1)-th singular triplets\\
$B_{\ell+1, \ell}$ & $\in \mathbf{R}^{(\ell+1) \times \ell}$, Bidiagonal matrix in ALB\\
$\hat{B}_{k+1}$ & Predicted of $B_{\ell+1}$ for (k+1)-th singular triplets\\

$U^{\bot}$ & A basis where $U$ is an orthogonal complement subspace\\
%GGD :
$\tilde{\mathcal{U}}$ & Denoised Image\\
$\mathcal{D}$ & Geodesic distance matrix\\
$\mathcal{G}$ & Gramian matrix\\
$\mathcal{I}$ & Original image \\
$I$ & Identity matrix\\
$\mathcal{U}$ & Noisy input image\\
$\Gamma$ & Shepard method's weights \\
$Q$ & Orthogonal matrix in randomized SVD \\

\hline %vectors :
$\bs{u}_i$ & $i$-th left singular vectors\\
$\bs{v}_i$ & $i$-th right singular vectors\\
$\bs{\nu}_l$ & $l$-th eigenvector \\

$\bs{w}_k$ & Vectors in MGSA\\
$\bs{c}_k$ & K columns which are randomly chosen in MGSA\\
$\bs{O}$ & Eigenvectors of the k-largest eigenvalues of $G$ in MCLA\\
%GGD :
$\tilde{\bs{u}}(\bs{x}_{ij})$ & Noise-free representation of the patch $\bs{u}(\bs{x}_{ij})$\\
$\bs{x}_{ij}$ & $ij$-th pixel of the image \\
$\bs{u}(\bs{x}_{ij})$ & Patch centered at the point $\bs{x}_{ij}$\\

\hline
$\mathcal{K}_\ell(A, b)$ & $= span\{\bs{b}, A\bs{b},...,A^{\ell-1}\bs{b}\}, \ell \geq 1$, Krylob subspace generated by $A$ and $\bs{b}$\\
%GGD :
$G(V,E)$ & A graph where $V$ denotes the vertex set and $E$ denotes the edge set\\
$\mathcal{N}(\bs{x}_{\tilde{k}})$ & Neighborhood around the pixel $\tilde{k}$ \\
\end{tabular}
\end{table}
%%%%%%%%%%%%%%%%%%%%%%%%%

%%%%%%%%%%%%%%%%%%%%%%%%%
\begin{table}[htp]
\caption {Abbreviations} \label{tab:abrv}
\begin{tabular}{p{2.2cm}|p{10cm}}
Abbreviation &  Description \\

\hline
%GGD : 
ALB & Augmented Lanczos Bidiagonalization\\
GGD & Geodesic Gramian Denoising \\
MCLA & Monte Carlo Low-Rank Approximation\\
PIME & Preconditioned Iterative Multimethod Eigensolver\\
PSNR & Peak Signal to Noise Ratio \\
RE & Relative Error \\
RSVD & Randomized Singular Value Decomposition\\
SSIM & Structural Similarity Index Measure \\
SVD & Singular Value Decomposition \\
\end{tabular}
\end{table}

%%%%%%%%%%%%%%%%%%%%%%%%%
%%%%%%%      Methods         %%%%%%%%%
%%%%%%%%%%%%%%%%%%%%%%%%%
\section{Method}\label{sec:method}
Here, first, we state the GGD-SVD algorithm with details and then introduce four diverse algorithms for approximating SVD. These four algorithms are Monte Carlo Rank Approximation (MCLA) \cite{friedland2006fast}, Augmented Lanczos Bidiagonalization \cite{baglama2005augmented}, Preconditioned Iterative Multimethod Eigensolver (PIME) \cite{wu2017primme_svds}, and Randomized Singular Value Decomposition (RSVD) \cite{halko2011finding}.

%%%%%%%%%%%%%%%%%%%%%%%%%
%%%%%%%          GGD           %%%%%%%%%
\subsection{Geodesic Gramian Denoising with Singular Value Decomposition}\label{sec:ggd}
GGD-SVD consists of five steps including one step that requires the computation of the SVD of a $n^2\times n^2$ data matrix where $n\times n$ is the size of the given matrix. The computational complexity of this step is $O(n^6)$ which we will replace with four singular vector approximation techniques.

In the first step, the input noisy image, denoted as $\mathcal{U}_{n\times n}$, is partitioned into overlapping square patches, denoted as $\bs{u}(\bs{x}_{ij})$'s; $i,j=1,\dots,n$, each with a fixed odd length, denoted as $\rho$, such that each patch is centered at each pixel of the image. For $k=n(i-1)+j$ and $1\le k \le n^2$, we write $\bs{u}(\bs{x}_{k})$ instead of $\bs{u}(\bs{x}_{ij})$ in some occasions for simplicity. Each patch $\bs{u}(\bs{x}_{k})$; $i=1,\dots,n^2$ can be represented as a $\rho^2$-dimensional point. A low-dimensional manifold underlies in this high-dimensional data cloud representing the set of patches of this image. In the second step, we create a graph structure, denoted as $G(V,E)$, using the neighbor search algorithm in Ref.~\cite{agarwal1999geometric}. Here, 1) the points $\{\bs{u}(\bs{x}_{k})\vert k=1,\dots,n^2\}$ are treated as vertices which are denoted by $V$; and, 2) the edge set $E$ is defined by the following process: for a given neighborhood parameter $\delta$, we join each vertex $\bs{u}(\bs{x}_{k})$ to its $\delta$ nearest neighbors such that the weight of the edge between the vertex $\bs{u}(\bs{x}_{k})$ and its neighbor vertex, denoted as $\bs{u}(\bs{x}_{k'})$, is the Euclidean distance between them, denoted as $d(k, k')$, where
\begin{equation}\label{eqn:eudist}
d(k, k')=\|\bs{u}(\bs{x}_{k})-\bs{u}(\bs{x}_{k'})\|_2.
\end{equation}
Then, the geodesic distance between two nodes on this network representing two patches is approximated by the shortest path distance between those nodes according to Floyd’s algorithm \cite{floyd1962algorithm}.

Third, the matrix of geodesic distances, denoted as $\mathcal{D} \in\mathbb{R}^{n^2\times n^2}_{\ge0}$, is transformed into a Gramian matrix, denoted as $\mathcal{G}_{n^2 \times n^2}$, using
\begin{equation}\label{eqn:gram}
\mathcal{G}[i,j]=-\big[\mathcal{D}[i,j]-\mu_i(\mathcal{D}) -\mu_j(\mathcal{D})+\mu(\mathcal{D})\big]\big/2,
\end{equation}
where $\mu_i(\mathcal{D})$ is the mean of the $i$-th row of the matrix $\mathcal{D}$, $\mu_j(\mathcal{D})$ is the mean of the $j$-th column of that matrix, and $\mu(\mathcal{D})$ is the mean of the full matrix \cite{lee2004nonlinear}. Fourth, GGD-SVD denoises the noisy patches $\bs{u}_{k}$ using only a few, denoted as $L$, prominent right singular vectors, defined as $\bs{\nu}_l$'s where $l=1, \dots, L$, of the Gramian matrix computed using Def.~\ref{def:svd} with $A=\mathcal{G}$ and $n_1=n_2=n$ because prominent singular vectors only represent image features excluding image noise. The noise-free representation of the noisy patch $\bs{u}_{k}$ is denoted as $\tilde{\bs{u}}_{k}$ which GGD-SVD produces by
\begin{equation}\label{eqn:denoise}
\tilde{\bs{u}}(\bs{x}_{k}) = \sum^{L}_{l=1} \langle \bs{u}(\bs{x}_{k}), \bs{\nu}_l \rangle \bs{\nu}_l,
\end{equation}
where the $l$-th prominent singular vector of the Gramian matrix is represented by $\bs{\nu}_l$ for $l=1,\dots,L$, and $\langle \cdot, \cdot \rangle$ represents the inner product operator. 

\begin{definition}\label{def:ed}
Consider, $Diag\left(\lambda_1, \dots, \lambda_n\right)$ denotes a diagonal matrix constructed with the vector $\left(\lambda_1, \dots, \lambda_n\right)$ as its diagonal. Let $A\in\mathbb{R}^{n\times n}$ be a matrix and $U_{n\times n}$ be a unitary matrix such that $U^TU=I$. Then, eigenvalue decomposition (ED) of $A$ is $A=U\Lambda U^T$, where $\Lambda_{n\times n}= $$Diag\left(\lambda_1, \dots \lambda_n\right)$, \cite{Golub1970}.
\end{definition}

\begin{definition}\label{def:svd}
Consider, $Diag\left(\sigma_1, \dots, \sigma_{\min(m,n)}\right)$ denotes a diagonal matrix constructed with the vector $\left(\sigma_1, \dots, \sigma_{\min(m,n)}\right)$ as its diagonal. Let $A\in\mathbb{R}^{m\times n}$ be a matrix, and $U_{m\times m}$ and $V_{n\times n}$ are unitary matrices such that $U^TU=I$ and $V^TV=I$, respectively. Then, the SVD of $A$ is $A=U\Sigma V^T$, where $\Sigma_{m\times n}= $$Diag\left(\sigma_1, \dots \sigma_{\min(m,n)}\right)$ with $\sigma_1\ge\dots\ge\sigma_{\min(m,n)}\ge0$. Here, for $l=1 ,\dots, n$, the column vector $\bs{\nu}_l$ represents $l$-th right singular vector of $A$ such that $V=[\bs{\nu}_1,\dots,\bs{\nu}_l,\dots,\bs{\nu}_{n}]$, \cite{Golub1970}. 
\end{definition}

In the fifth step, GGD-SVD constructs the noise-free image from the denoised patches. For that, it estimates each $k$-th pixel's intensity of the image by using the pixels of $\rho^2$ overlapping patches coincident with this $k$-th pixel. Thus, for the sake of convenient interpretation, we introduce a new index $t_n$ for each pixel $\bs{x}_t \in \mathcal{N}(\bs{x}_k)$, where
\begin{equation}
\mathcal{N}(\bs{x}_{k})=\{\bs{x}_{t} \ \vert \ \ \|\bs{x}_{k}-\bs{x}_{t}\|_{\infty}\le \rho/2\},
\end{equation}
such that the physical location of the pixel of the patch $\tilde{\bs{u}}(\bs{x}_t)$ at the new index, denoted as $[\tilde{\bs{u}}(\bs{x}_t)]_{t_n}$, is the same physical location of the pixel $\bs{x}_k$. Then, all these estimations are combined using a moving least square approximation according to Shepard’s method \cite{Shepard1968}, as
\begin{equation}\label{eqn:merge}
\tilde{\mathcal{U}}(\bs{x}_{k})=\sum_{\bs{x}_t \in \mathcal{N}(\bs{x}_{k})} \Gamma (\bs{x}_{k},\bs{x}_t) [\tilde{\bs{u}}(\bs{x}_t)]_{t_n},
\end{equation}
\cite{gajamannage2020patch}, where the weights are given by
\begin{equation}\label{eqn:shepard_weight}
\Gamma(\bs{x}_{k},\bs{x}_t)=\frac{e^{-\|\bs{x}_{k}-\bs{x}_t\|^2}}{\sum_{\bs{x}_{t'}\in \mathcal{N}(\bs{x}_{k})} e^{-\|\bs{x}_{k}-\bs{x}_{t'}\|^2}}.
\end{equation}
This weighting term weights nearby pixels with higher weight whereas faraway pixels with lower weight. Thus, Eqn.~\eqref{eqn:merge} assures that the pixel $\bs{x}_{k}$ of the denoised image is heavily influenced by the pixels at the same location of the nearby patches. GGD-SVD algorithm is summarized in Alg.~\ref{alg:ggd}.

\begin{algorithm}[H]
\caption{ 
Geodesic Gramian Denoising with SVD (GGD-SVD).\\
Inputs:  noise-contaminated image ($\mathcal{U}_{n\times n}$), length of a square patch ($\rho$),  size of the nearest neighborhood ($\delta$), and threshold for the singular vectors ($L$).\\
Outputs:  denoised image ($\tilde{\mathcal{U}}_{n\times n}$).}
\begin{algorithmic}[1]
\State  Produce a set of $n^2$ overlapping $\rho \times \rho$ patches of the noisy image $\mathcal{U}_{n\times n}$ and denote this set by $\{\bs{u}(\bs{x}_k)\vert k=1,\dots,n^2\}$.	
\State Utilize the nearest neighbor search algorithm in Ref.~\cite{agarwal1999geometric} to construct the graph $G(V,E)$ of the patch set. Produce the geodesic distance matrix $\mathcal{D}$ by approximating the geodesic distances on this graph using Floyd's algorithm in Ref.~\cite{floyd1962algorithm}.	
\State Use Eqn.~\eqref{eqn:gram} to construct the Gramian matrix $\mathcal{G}$ from $\mathcal{D}$ .	
\State Decompose the right-sided singular vectors $\left\{\bs{\nu}_l\vert l = 1,\dots L\right\}$ of the $L$ biggest eigenvalues of the Gramian matrix $\mathcal{G}$ using SVD.
\State Construct the denoised patches $\{\tilde{\bs{u}}(\bs{x}_k)\vert k=1,\dots,n^2\}$ using the right singular vectors $\left\{\bs{\nu}_l\vert l = 1,\dots L\right\}$ according to Eqn.~\eqref{eqn:denoise}.	
\State Use Eqns.~\eqref{eqn:merge} and \eqref{eqn:shepard_weight} to merge denoised patches and construct the noise-free image $\tilde{\mathcal{U}}_{n\times n}$.
\end{algorithmic}\label{alg:ggd}
\end{algorithm}

For an input image of size $n\times n$, step 4 of Alg.~\ref{alg:ggd} requires to compute the ED [see the computation of $\{\bs{\nu}_l\vert l = 1,\dots L\}$ of Eqn.~\eqref{eqn:denoise}] of a Gramian matrix $\mathcal{G}$ of size $n^2\times n^2$. In GGD-SVD, we employ regular SVD that calculates the entire singular vector spectrum of $\mathcal{G}$ using Def.~\ref{def:svd} and then we choose $L$ prominent singular vectors for denoising. Especially, the implementation of SVD is done by two processes: first, the given matrix of size $m \times n$ is reduced to a bidiagonal matrix form using Householder transformations with $\mathcal{O}(mn \min(m, n))$ operations, and then use the QR algorithm\footnote{QR algorithm is a procedure to calculate the eigenvalues and singular vectors of a given matrix using a Schur decomposition.} in Ref.~\cite{Francis1961, Francis1962}, to compute the SVD of the bidiagonal matrix with $\mathcal{O}(n)$ operations \cite{Golub1970}. Thus, the computational complexity of GGD-SVD is dominated by step 4 of Alg.~\ref{alg:ggd} with $\mathcal{O}(n^6)$ as it computed SVD of Gramian matrix of the size $n^2 \times n^2$ for an input image of $n\times n$. Thus, in the Secs.~\ref{sec:mcla}-\ref{sec:rsvd}, we present four singular vector approximation schemes that we can use to approximate the required number of prominent SVs and then replace the SVD-SVD of step 4 of Alg.~\ref{alg:ggd} to increase the computational efficiency of GGD-SVD.

%%%%%%%%%%%%%%%%%%%%%%%%%
%%%%%%%          MCAL           %%%%%%%%%
\subsection{Monte Carlo Low-Rank Approximation}\label{sec:mcla} %Julia
Monte Carlo Low-Rank Approximation (MCLA) \cite{friedland2006fast}, uses an iterative Monte Carlo sampling approach to approximate the SVD of a data matrix $A$ of size $m\times n$ by computing the $L$-rank approximation, for some $L\in\mathbb{N}$ where $L<m,n$. Each iteration involves the reading of $\mathcal{O}(L)$ of columns or rows of the data matrix that eventually attains a computational time complexity of $\mathcal{O}(Lmn)$. Note that, since the Gramian matrix $\mathcal{G}$ of GGD-SVD is of size $n^2\times n^2$ for a given image of size $n\times n$, the computational time complexity of MCLA generating $L$-prominent singular vectors is  $\mathcal{O}(Ln^4)$. Instead of precisely computing the singular vectors of $\mathcal{G}$ as stated in Step 4 of Alg.~\ref{alg:ggd}, we approximate those singular vectors using MCLA and then perform the denoising, in which this hybrid framework is denoted as GGD-MCLA.

We formulate MCLA for a general data matrix $A$ of size $m\times n$, and then we replace this $A_{m\times n}$ with $\mathcal{G}_{n^2\times n^2}$. First, we start the MCLA algorithm by computing the initial $L$-rank approximation, denoted as $B_0\in \mathbb{R}^{m\times n}$, of $\mathcal{G}$. For that, let $\bs{c}^{(0)}_1,\dots, \bs{c}^{(0)}_n \in \mathbb{R}^{m}$ be the $n$ columns of $A$. Choose $L$ integers $1\le n_1<\dots,n_L\le n$ randomly and let $\bs{\nu}^{(0)}_1,\dots,\bs{x}^{(0)}_L\in\mathbb{R}^m$ be the orthonormal set obtained from $\bs{c}^{(0)}_{n_1},\dots,\bs{c}^{(0)}_{n_L}$ using the modified Gram-Schmidt algorithm given in Alg.~5.2.6 of Ref.~\cite{golub2013matrix}. Then, set
\begin{eqnarray}\label{eqn:B0}
B^{(0)}=\sum^L_{l=1}\bs{x}^{(0)}_l\left(A^T\bs{x}^{(0)}_l\right)^T.
\end{eqnarray}

%%%%%%%%%%%%
\begin{algorithm}[H]
\caption{ 
Monte Carlo Low-Rank Approximation (MCLA) \cite{friedland2006fast}.\\
Inputs:  data matrix $\left(A\in \mathbb{R}^{m\times n}\right)$, rank ($k$), number of iterations $\left(T_I\right)$, tolerance ($\eta$).\\
Outputs:  rank $L$ approximation $B\in \mathbb{R}^{m\times n}$ of $A$, approximated $L$ prominent singular values $\left(\{\sigma_l \ \vert \ l = 1, \dots, L\}\right)$, and their approximated right singular vectors $\left(\{\bs{\nu}_l \ \vert \ l = 1, \dots, L\}\right)$
}
\begin{algorithmic}[1]
\State  Compute the $L$-rank approximation of $A$ with $L$ columns (or rows) of $A$ using Eqn.\eqref{eqn:B0}.
\For{$t = 1$ to $T_I$}	
\State Select $L'$ columns (or rows), $\bs{c}^{(t)}_1, \dots, \bs{c}^{(t)}_{L'}\in \mathbb{R}^m$, from $A$ at random. Consider $\left\{\bs{x}^{(t-1)}_l \in \mathbb{R}^m\ \Big\vert \ l=1,\dots, L\right\}$ is the orthonormal set involved in the computation of $B^{(t-1)}$. 
\State For $L\le p\le L+L'$, compute the orthonormal basis $\bs{w}^{(t)}_1,\dots,\bs{w}^{(t)}_L, \dots, \bs{w}^{(t)}_p$ of $L'$ columns and orthonormal set and compute $C$ using the orthonormal basis as in Eqn.~\eqref{eqn:c}.
\State Compute $L$-prominent eigenvectors $\left\{\tilde{\bs{x}}^{(t)}_l \in \mathbb{R}^p\ \Big\vert \ l=1,\dots, L\right\}$ of $C$ in Eqn.~\eqref{eqn:c}. Convert these eigenvectors to $\left\{\bs{x}^{(t-1)}_l \in \mathbb{R}^m\ \Big\vert \ l=1,\dots,L\right\}$ using Eqn.~\eqref{eqn:order}
\State Compute $B^{(t)}$ using Eqn.~\eqref{eqn:Bt}. 
\If{$\|B^{(t-1)}\|/\|B^{(t)}\|>1-\eta$}
\State Goto 11.
\EndIf
\EndFor
\State $B:=B^{(t)}$; approximated singular values of $A$ are $\left\{\sigma_l \ \Big\vert \ l=1, \dots, L\right\}$ where $\sigma_l:=\sqrt{\left(A^T\bs{x}^{(t)}_i\right)^T\left(A^T\bs{x}^{(t)}_i\right)}$; and
approximated singular vectors of $A$ are $\left\{\bs{\nu}^{(t)}_l=\frac{1}{\sigma_l}A^T\bs{x}^{(t)}_l\in \mathbb{R}^m \ \Big\vert \ l=1, \dots, L\right\}$.
\end{algorithmic}\label{alg:mcla}
\end{algorithm}
%%%%%%%%%%%%

Now, for $t=0, \dots, T$, we compute the $L$-rank approximation of $A$, denoted as $B^{(t)}\in \mathbb{R}^{m\times n}$, which is improved from $B^{(t-1)}\in \mathbb{R}^{m\times n}$. Let $\bs{x}^{(t-1)}_1,\dots,\bs{x}^{(t-1)}_L\in\mathbb{R}^m$ is the orthonormal set involved in the computation of $B^{(t-1)}$ such that $B^{(t-1)}=\sum^L_{l=1}\bs{x}^{(t-1)}_l\left(A^T\bs{x}^{(t-1)}_l\right)^T$. Now, we choose random $L'$ columns of $A$ where $1\le L' \le n$ and denote them by $\bs{c}^{(t)}_1, \dots, \bs{c}^{(t)}_{L'}\in\mathbb{R}^m$. Let, $X$ be the subset spanned by $\bs{x}^{(t-1)}_1,\dots,\bs{x}^{(t-1)}_L, \bs{c}^{(t)}_1, \dots, \bs{c}^{(t)}_{L'}$. For $L<p\le L+L'$, we generate an orthonormal basis $\bs{w}^{(t)}_1,\dots,\bs{w}^{(t)}_L, \dots, \bs{w}^{(t)}_p\in\mathbb{R}^m$ of $X$ obtained from $\bs{x}^{(t-1)}_1,\dots,\bs{x}^{(t-1)}_L, \bs{c}^{(t)}_1, \dots, \bs{c}^{(t)}_l$. For $L<p\le L+L'$ using a modified Gram-Schmidt algorithm. Now, define a $p\times p$ non-negative definite matrix \begin{eqnarray}\label{eqn:c}
C=\biggl(\left(A^T\bs{w}^{(t-1)}_i\right)^T\left(A^T\bs{w}^{(t-1)}_i\right)\biggr)^p_{i,j=1}.
\end{eqnarray}
We compute $L$-prominent eigenvectors of $C$ that we denote as $\tilde{\bs{x}}^{(t)}_1,\dots,\tilde{\bs{x}}^{(t)}_L\in \mathbb{R}^p$. We compute $\left\{\bs{x}^{(t-1)}_l \in \mathbb{R}^m\ \Big\vert \ l=1,\dots,L\right\}$ such that 
\begin{eqnarray}\label{eqn:order}
\left(\bs{x}^{(t)}_1, \dots, \bs{x}^{(t)}_L\right)
=\left(\bs{w}^{(t)}_1, \dots, \bs{w}^{(t)}_p\right)
\left(\tilde{\bs{x}}^{(t)}_1, \dots, \tilde{\bs{x}}^{(t)}_L\right)
\end{eqnarray}
Then, 
\begin{eqnarray}\label{eqn:Bt}
B^{(t)}=\sum^L_{l=1}\bs{x}^{(t)}_l\left(A^T\bs{x}^{(t)}_l\right)^T,
\end{eqnarray}
is the improved rank $L$ approximation of $A$ from $B^{(t-1)}$. Now, while the approximated $L$ prominent singular values of $A$ are given by  
\begin{eqnarray}
\sigma_l=\sqrt{\left(A^T\bs{x}^{(t)}_i\right)^T\left(A^T\bs{x}^{(t)}_i\right)},
\end{eqnarray}
corresponding right singular values, denoted by $\left\{\bs{\nu}^{(t)}_l\Big\vert l = 1, \dots, L\right\}$, are given by
\begin{eqnarray}\label{eqn:rsv}
\bs{\nu}^{(t)}_l=\frac{1}{\sigma_l}A^T\bs{x}^{(t)}_l,
\end{eqnarray}

For a given image of size $n\times n$, we compute the Gramian matrix $\mathcal{G}_{n\times n}$ using Steps 1-3 in Alg.~\ref{alg:ggd}. Then, we approximate $L$ prominent singular vectors using Alg.~\ref{alg:mcla} with $A=\mathcal{G}_{n\times n}$. Finally, we use those singular vectors to compute the noise-free image as explained in Steps 5-6 of Alg.~\ref{alg:ggd}.

%%%%%%%%%%%%%%%%%%%%%%%%%
%%%%%%%          ALB           %%%%%%%%%
\subsection{Augmented Lanczos Bidiagonalization}\label{sec:alb}
Augmented Lanczos Bidiagonalization (ALB) \cite{baglama2005augmented}, inputs a matrix of order $n\times m$, denoted as $A$ , and  computes a few largest (or smallest) singular triplets of it. ALB performs a sequence of projections of $A$ onto a carefully chosen low-dimensional Krylov subspaces. ALB is implemented as an augmentation process of Krylov subspaces where those subspaces are revealed similar to the regular Lanczos bidiagonalization technique in which Ritz vectors or harmonic Ritz vectors impose the augmentation. The hybrid framework replacing the explicit singular vector computation of $\mathcal{G}$, i.e., Step 4 of Alg.~\ref{alg:ggd}, by the approximation of singular vectors using ALB, is denoted as GGD-ALB.

The first step of ALB is to approximate the data matrix $A\in\mathbb{R}^{m\times n}$ by a bidiagonal matrix, denoted as  $B\in\mathbb{R}^{m\times n}$, using the Lanczos bidiagonalization process \cite{baglama2005augmented}. Bidiagonalization is a widely used kernel that transforms a full matrix into a bidiagonal form using orthogonal transformations. The bidiagonalization procedure is a preprocessing step that significantly lowers the cost of the implicit implementation of the SVD algorithm. For given data matrix $A$ and the initial unit vector $\bs{p}^{(1)}\in\mathbb{R}^n$, this process yields the decomposition
\begin{eqnarray}\label{eqn:LB1}
AP^{(t)}=Q^{(t)}B^{(t)},
\end{eqnarray}
\begin{eqnarray}\label{eqn:LB2}
A^TQ^{(t)}=P^{(t)}\left(B^{(t)}\right)^T+\bs{r}^{(t)}\left(\bs{e}^{(t)}\right)^T,
\end{eqnarray}
where $P^{(t)}\in\mathbb{R}^{n\times h}$, $Q^{(t)}\in\mathbb{R}^{m\times h}$, $\left(P^{(t)}\right)^TP^{(t)}=I^{(t)}$, $P^{(t)}\bs{e}^{(1)}=\bs{p}^{(1)}$, $\left(Q^{(t)}\right)^TQ^{(t)}=I^{(t)}$, $\bs{r}^{(t)}\in\mathbb{R}^n$, and $\left(P^{(t)}\right)^T\bs{r}^{(t)}=0$.  Lanczos bidiagonalization governing Eqns. \eqref{eqn:LB1} and \eqref{eqn:LB2} are numerically implemented in Alg.~\ref{alg:gkb}. It is not necessary to be columns of $P$ and $Q$ are orthogonal if they are not reorthogonalized since finite precision arithmetic is performed in Alg.~\ref{alg:gkb}.

\begin{algorithm}[H]
\caption{Lanczos bidiagonalization \cite{baglama2005augmented}.\\
Input: data matrix ($A\in \mathbb{R}^{m\times n}$), initial unit-length vector ($\bs{p}_1\in\mathbb{R}^n$), and the number of bidiagonalization steps ($T_B$). \\
Output: upper bidiagonal matrix ($B\in\mathbb{R}^{h\times h}$) with entries $\alpha_j$ and $\beta_j$,
matrices of orthonormal columns ($P=[\bs{p}_1,\dots, \bs{p}_n] \in\mathbb{R}^{n\times h}$ and  $Q=[\bs{q}_1,\dots, \bs{q}_n] \in\mathbb{R}^{m\times h}$), and residual error ($\bs{r}\in \mathbb{R}^n$). }\label{alg:gkb}
\begin{algorithmic}[1]
\State Initialization: $P_1=\bs{p}_1$, $\bs{q}_1=A\bs{p}_1$, $\alpha_1=\|\bs{q}_1\|_2$, $\bs{q}_1=\bs{q}_1/\alpha_1$, and $Q_1=\bs{q}_1$
\For {$t=1$ to $T_B$ }     
    \State $\bs{r}^{(t)}=A^T\bs{q}^{(t)}-\alpha^{(t)}\bs{p}^{(t)}$		
    \State Reorthogonalization: $\bs{r}^{(t)} = \bs{r}^{(t)} - P^{(t)}\left(\left(P^{(t)}\right)^T\bs{r}^{(t)}\right)$    
    \If {$t<T_B$}
   	\State $\beta^{(t)}=\|\bs{r}^{(t)}\|_2$; $\bs{p}^{(t+1)}=\bs{r}^{(t)}/\beta^{(t)}$, and $P^{(t+1)}=[P^{(t)},\bs{p}^{(t+1)}]$
   	\State $\bs{q}^{(t+1)}=A\bs{p}^{(t+1)}-\beta^{(t)}\bs{q}^{(t)}$
   	\State Reorthogonalization: $\bs{q}^{(t+1)} = \bs{q}^{(t+1)} - Q^{(t)}\left(\left(Q^{(t)}\right)^T\bs{q}^{(t+1)}\right)$    
   	\State $\alpha^{(t+1)}=\|\bs{q}^{(t+1)}\|_2$; $\bs{q}^{(t+1)}=\bs{q}^{(t)}/\alpha^{(t+1)}$, and $Q^{(t+1)}=[Q^{(t)},\bs{q}^{(t+1)}]$
    \EndIf    
\EndFor
\State $B=B^{(t)}$, $P=P^{(t)}$, and $Q=Q^{(t)}$
\end{algorithmic}
\end{algorithm}

Now, we perform SVD on this bidiagonal matrix ($B_{h\times h}$) according to Def.~\ref{def:svd} and generate singular triplets $\{\sigma_l, \bs{u}_l, \bs{v}_l\vert l = 1, \dots, h\}$ where $\sigma_1\ge\dots\ge\sigma_h\ge0$. Entries of the singular triplet hold the relation
\begin{eqnarray}\label{eqn:svd-trip}
B\bs{v}_l=\sigma_l\bs{u}_l, \ \ \ B^T\bs{u}_l=\sigma_l\bs{v}_l, \ \ \ l=1, \dots, h.
\end{eqnarray}
 We determine approximate singular triplets $\left\{\tilde{\sigma_l},\tilde{\bs{u}}_l, \tilde{\bs{u}}_l\right\}$, $l=1, \dots, h$ of $A$ from singular triplets of $B$ by
\begin{eqnarray}\label{eqn:svd-B}
\tilde{\sigma}_l=\sigma_l, \ \ \ \tilde{\bs{u}}_l=Q\bs{u}_l, \ \ \ \tilde{\bs{v}}_l=P\bs{v}_l
\end{eqnarray}
Combining Eqn.~\eqref{eqn:svd-trip} with Eqns.~\eqref{eqn:LB1} and ~\eqref{eqn:LB2} gives  
\begin{eqnarray}\label{eqn:LB3}
A\tilde{\bs{v}}_l=\tilde{\sigma}_l\tilde{\bs{u}}_l, \ \ \ A^T\tilde{\bs{u}}_l=\tilde{\sigma}_l\tilde{\bs{v}}_l+\bs{r}\bs{e}^T\bs{u}_j
\end{eqnarray}
Eqn.~\eqref{eqn:LB3} suggests that the singular triplets $\{\tilde{\sigma}_l, \tilde{\bs{u}}_l, \tilde{\bs{v}}_l\vert l = 1, \dots, h\}$ be qualified to be approximated singular triplets of $A$ if $\bs{r}\bs{e}^T\bs{u}_j$ is sufficiently small. Especially, the ALM scheme accepts them as singular triplets if
\begin{eqnarray}\label{eqn:rtip-cond}
\|\bs{r}\|_2\vert\bs{e}^T\bs{u}_l\vert\le\delta\|A\|_2,
\end{eqnarray}
where $\delta\in\mathbb{R}^+$ and $\|A\|_2$ is approximated as the largest SV of $B$.

Numerical instability is considered to be an issue pertaining to the implicitly restarted Lanczos bidiagonalization technique since its propagated round-off errors delay or prevent convergence of eigenpairs \cite{baglama2005augmented}. We augment Krylov subspaces by some Ritz vectors as presented in Ref.~\cite{Sorensen1992} to increase this numerical stability. Consider that the approximated right singular vectors $\{\tilde{\bs{v}}_l, l=1,\dots, h\}$ of $A$ in Eqn.~\eqref{eqn:svd-B} are Ritz vectors of $A^TA$ associated with Ritz values $\{\tilde{\sigma}^2_l, l=1,\dots, h\}$. Let, the Ritz vectors $\{\tilde{\bs{v}}_l, l=1,\dots, L\}$ for the L-largest Ritz values be available and $\bs{r}\neq 0$. We introduce the matrix
\begin{eqnarray}\label{eqn:ritzP}
\tilde{P}_{L+1}=\left[\tilde{\bs{v}}_1, \dots, \tilde{\bs{v}}_L,\bs{p}_{h+1}\right].
\end{eqnarray}
Let $\tilde{\bs{r}}_L$ be the remainder orthogonal to the vectors $\{\bs{u}_l, l=1, \dots, h\}$ defined in Ref.~\cite{baglama2005augmented} and introduce the matrices
\begin{eqnarray}\label{eqn:ritzQ}
\tilde{Q}_{L+1}=\left[\tilde{\bs{u}}_1, \dots, \tilde{\bs{u}}_L,\frac{\tilde{\bs{r}}_{L}}{\|\tilde{\bs{r}}_{L}\|}\right]
\end{eqnarray}
and 
\begin{equation}\label{eqn:ritzB}
\tilde{B}_{L+1} = 
\begin{bmatrix}
\tilde{\sigma}_1  \ \ \ & 0 & \tilde{\rho}_1\\
\\
\ \ \  & \tilde{\sigma}_L & \tilde{\rho}_L\\
0  \ \ \ \ & & \tilde{\alpha}_{L+1}\\
\end{bmatrix}
\in \mathbb{R}^{(L+1)\times (L+1)},
\end{equation}
where $\tilde{\rho}_l=(\tilde{\bs{u}}_l)^TA\bs{p}_{h+1}, l=1, \dots, h$. Then, an analogous representation of Eqns.~\eqref{eqn:LB1} and \eqref{eqn:LB2} using Eqns.~\eqref{eqn:ritzP}, \eqref{eqn:ritzQ}, and \eqref{eqn:ritzB} are
\begin{eqnarray}\label{eqn:ritz_LB1}
A\tilde{P}_{L+1}=\tilde{Q}_{L+1}\tilde{B}_{L+1},
\end{eqnarray}
and 
\begin{eqnarray}\label{eqn:ritz_LB2}
A^T\tilde{Q}_{L+1}=\tilde{P}_{L+1}\left(\tilde{B}_{L+1}\right)^T+\bs{r}_{L+1}\left(\bs{e}_{L+1}\right)^T,
\end{eqnarray}
respectively.

The above augmentation by Ritz vectors offers accurate approximations for the largest singular triplets of $A$. However, Ref.~\cite{Kokiopoulou2004} states that the augmentation by harmonic Ritz vectors may offer faster convergence than that by Ritz vectors when the focus is to reveal the smallest singular triplets of $A$. Let, the $L'$ smallest singular triplets of $B_{h\times h+1}$ are presented by the matrices,
\begin{equation}
\begin{split}\label{eqn:ritz_SVD}
U' = [\bs{u}'_1, \dots, \bs{u}'_{L'}]\in \mathbb{R}^{h\times L'},\\
V' = [\bs{v}'_1, \dots, \bs{v}'_{L'}]\in \mathbb{R}^{(h+1)\times L'},\\
\Sigma' = Diag[\sigma'_1, \dots, \sigma'_{L'}]\in \mathbb{R}^{L'\times L'}.\\
\end{split}
\end{equation}
We now derive the relations analogous to Eqns.~\eqref{eqn:LB1} and \eqref{eqn:LB2} for harmonic Ritz vectors. We introduce QR factorization 
\begin{equation}\label{eqn:ritzQR}
\begin{bmatrix}
B_h^{-1}U^{'}_L\Sigma^{'}_L & -\beta_h B_h^{-1}\bs{e}_h\\
0 & 1\\
\end{bmatrix}
=Q'_{L+1}R'_{L+1},
\end{equation}
where columns of $Q'_{L+1}\in \mathbb{R}^{(h+1)\times(L+1)}$ are orthonormal and $R'_{L+1}\in \mathbb{R}^{(L+1)\times(L+1)}$ is upper triangular. We denote the matrices
\begin{eqnarray}\label{eqn:hRitzP}
\hat{P}_{L+1}=\left[\hat{\bs{p}}_1, \dots, \hat{\bs{p}}_{L+1}\right]=P_{h+1}Q'_{L+1} 
\end{eqnarray}
\begin{eqnarray}\label{eqn:hRitzQ}
\hat{Q}_{L+1}=\left[\hat{Q}_L, \hat{\bs{q}}_{L+1}\right]\in\mathbb{R}^{m\times (L+1)}, \ \ \text{with} \ \ \hat{Q}_L=Q_hU'_L,
\end{eqnarray}
and
\begin{equation}\label{eqn:hRitzB}
\tilde{B}_{L+1} = 
\begin{bmatrix}
\sigma'_1  \ \ \ & 0 & \tilde{\gamma}_1\\
\\
\ \ \  & \sigma'_L & \hat{\gamma}_L\\
0  \ \ \ \ & & \hat{\alpha}_{L+1}\\
\end{bmatrix}
\left(R'_{L+1}\right)^{-1}\in \mathbb{R}^{(L+1)\times (L+1)},
\end{equation}
where $\hat{\gamma}_l=-\beta_h(\hat{\bs{q}}_l)^T\hat{\bs{q}}_h+(\hat{\bs{q}}_l)^TA\bs{p}_{h+1}, l=1, \dots, L$.

 Then, an analogous representation of Eqns.~\eqref{eqn:LB1} and \eqref{eqn:LB2} using Eqns.~\eqref{eqn:hRitzP}, \eqref{eqn:hRitzQ}, and \eqref{eqn:hRitzB} are
\begin{eqnarray}\label{eqn:ritz_LB3}
A\hat{P}_{L+1}=\hat{Q}_{L+1}\hat{B}_{L+1},
\end{eqnarray}
and 
\begin{eqnarray}\label{eqn:ritz_LB4}
A^T\hat{Q}_{L+1}=\hat{P}_{L+1}\left(\hat{B}_{L+1}\right)^T+\bs{r}_{L+1}\left(\bs{e}_{L+1}\right)^T.
\end{eqnarray}
respectively.

%%%%%%%%%%%%
\begin{algorithm}[H]
\caption{Augmented Lanczos Bidiagonalization (ALB).\\
Input: data matrix ($A\in \mathbb{R}^{m\times n}$), initial unit-length vector ($\bs{p}_1\in\mathbb{R}^n$), amount of bidiagonalization steps ($T_B$), amount of required singular triplets ($L$), tolerance ($\delta$), machine epsilon ($\epsilon$), and Boolean variable harmonic ($h$). \\Output: approximated singular triplets $\left(\left\{\sigma_l, \bs{u}_l,\bs{v}_l\right\}_{l=1}^L\right)$.}\label{alg:aug}

\begin{algorithmic}[1]
\State Produce the bidiagonal matrix $B$ by running the Lanczos bidiagonalization Algorithm, e.i., Alg.~\ref{alg:gkb}, on $A$.
\State Compute the SVD of $B$ using Def.~\ref{def:svd} and generate the singular triplets $\{\sigma_l, \bs{u}_l, \bs{v}_l\vert l = 1, \dots, h\}$ where $\sigma_1\ge\dots\ge\sigma_m\ge0$.

\vspace{5mm}
\item[] Checking the convergence:
\If{all $l=1,\dots,L$ singular triplets satisfy the inequality in Eqn.~\eqref{eqn:rtip-cond}}
\State Exit
\EndIf

\vspace{5mm}
\item[] Computation of the augmentation vectors:
   \If{not harmonic vectors or $\sigma_h/\sigma_1>\epsilon^{-1/2}$}
      \State Define $P=\tilde{P}_{L+1}$, $Q=\tilde{Q}_{L+1}$, and $B=\tilde{B}_{L+1}$ where $\tilde{P}_{L+1}$, $\tilde{Q}_{L+1}$, and $\tilde{B}_{L+1}$ are given in Eqns.~\ref{eqn:ritzP}, \eqref{eqn:ritzQ}, and \ref{eqn:ritzB}, respectively.
      \EndIf
      \If{harmonic and $\sigma_h/\sigma_1\le\epsilon^{-1/2}$}
      \State Compute the partial SVD of $B_{h\times h+1}$ using Eqn.~\eqref{eqn:ritz_SVD} and QR-factorization using Eqn.~\eqref{eqn:ritzQR}. Determine matrices $P=\hat{P}_{L+1}$, $Q=\hat{Q}_{L+1}$, and $B=\hat{B}_{L+1}$ given by Eqns.~\eqref{eqn:hRitzP}, \eqref{eqn:hRitzQ}, and \eqref{eqn:hRitzB}, respectively.
       \EndIf
       
\vspace{5mm}       
\State The matrices $P$, $Q$, and $B$, as well as the vector $r$ satisfy $$AP = QB \ \  ; \ \ A^TQ = PB^T + \bs{r}\bs{e}_L^T.$$ 
Append both $h-L$ columns into the matrices $P$ and $Q$, and $h-L$ rows and columns into the matrix $B$. Denote the aforesaid matrices $P$, $Q$, and $B$ by $P_h$, $Q_h$, and $B_h$, respectively. Compute the new residual vector and denote it by $\bs{r}_h$.
\State Goto 4.
\end{algorithmic}
\end{algorithm}
%%%%%%%%%%%%

For an image of size $n\times n$, we compute the Gramian matrix $\mathcal{G}_{n\times n}$ using Steps 1-3 in Alg.~\ref{alg:ggd}. Then, we approximate $L$ prominent singular vectors using Alg.~\ref{alg:aug} with $A=\mathcal{G}_{n\times n}$. Finally, we use those singular vectors to compute the noise-free image as explained in Steps 5-6 of Alg.~\ref{alg:ggd}.

%%%%%%%%%%%%%%%%%%%%%%%%% 
%%%%%%%          PIME           %%%%%%%%%
\subsection{Preconditioned Iterative Multimethod Eigensolver}\label{sec:pime} 
The singular vector approximation scheme that we are utilizing here is named Preconditioned Iterative Multimethod Eigensolver (PIME) \cite{wu2017primme_svds}, which is an extension of the state-of-the-art package called PRIMME \cite{stathopoulos2010primme}. PRIMME is capable of solving a range of both large and sparse singular value problems either with or without preconditioning; thus, PRIMME assures unprecedented efficiency, robustness, and accuracy, for both the smallest and the largest singular triplets. We denote the hybrid framework, replacing the explicit singular vector computation of $\mathcal{G}$, i.e., Step 4 of Alg.~\ref{alg:ggd}, using the approximation of singular vectors using PIME, by GGD-PIME.

PIME inputs a data matrix, denoted as $A\in\mathbb{R}^{m\times n}$, and two parameters, namely, the number of targeted singular triplets, denoted as $T_I$, and tolerance, denoted as $\delta$, which then outputs approximated $L$-many singular triplets. PIME depends on eigensolvers that rely on an equivalent eigenvalue formulation on $C$ and $B$, where
\begin{equation}\label{eq:C}
C=A^TA\in \mathbb{R}^{n\times n} \  \ (\text{or} \ \ C=AA^T\in \mathbb{R}^{m\times m}),
\end{equation}
and
\begin{equation}\label{eq:B}
B =
\begin{bmatrix}
0 & A^T\\
A & 0 
\end{bmatrix}\mathbb{R}^{(m+n) \times (m+n)}. 
\end{equation}
Without preconditioning the convergence with respect to iterations, starting PIME on $C$ is much faster than starting on $B$. Specifically, the cost per iteration for starting on $C$ is up to two times cheaper than that on $B$ (since the two dimensions, $n$ versus $n+m$). PIME works on $C$ during the first stage of the technique and switches to the second stage where it works on $B$ if further accuracy is required.

First, PIME computes eigenpairs of $C$ in Eqn.~\eqref{eq:C} using Def.~\ref{def:ed}. Let, $\left(\lambda^C_i, \bs{x}^C_i\in\mathbb{R}^n\right), i = 1, \dots, n$ be eigenpairs of $C$ computed using Def.~\ref{def:ed}. Then, Rayleigh-Ritz scheme is used to approximate the eigenpairs, denoted as $\left(\tilde{\lambda}^C_i, \tilde{\bs{x}}^C_i\right), i = 1, \dots, n$, on the vector basis of $\left[\bs{x}^C_1, \dots, \bs{x}^C_n\right]$. Set, the singular triplets $\left(\tilde{\sigma}_i, \tilde{\bs{u}}_i, \tilde{\bs{v}}_i\right), i = 1, \dots, n$ such that $\tilde{\sigma}_i = \sqrt{\vert\tilde{\lambda}^C_i\vert}$,  $\tilde{\bs{v}}_i = \tilde{\bs{x}}^C_i$, and $\tilde{\bs{u}}_i = A \tilde{\bs{x}}^C_i\tilde{\sigma}_i$. The stopping criterion that imposes the residual norm of the singular value triplets of $A$ is less than $\|A\|_2\delta$, i.e.,
\begin{equation}\label{eq:convergeA}
\|\tilde{\bs{r}}_i\|=\sqrt{\|A\tilde{\bs{v}}_i-\tilde{\sigma}_i\tilde{\bs{u}}_i\|_2+\|A\tilde{\bs{u}}_i-\tilde{\sigma}_i\tilde{\bs{v}}_i\|_2}<\|A\|_2\delta
\end{equation}
is set in PIME to ensure the convergence. Eqn.~\eqref{eq:convergeA} can also be transformed into a criterion that ensures the convergence  of the eigensolver on $C$ as
\begin{equation}\label{eq:convergeC}
\|\tilde{\bs{r}}^C_i\|=\|C\tilde{\bs{x}}^C_i-\tilde{\lambda}^C_i\tilde{\bs{x}}_i\|_2<\sqrt{\vert\tilde{\lambda}^C_i\vert\|C\|_2\delta}
\end{equation}
This eigensolver is implemented until the convergence criterion is satisfied by all the requested triplets. However, it is plausible for this eigensolver to reach its best accuracy before the residual norm falls below the aforementioned convergence limit. While the stagnation of the eigensolver is observed if the tolerance is set below that limit, the execution time of the second state may be increased if the tolerance is overestimated.

In the second stage, PIME computes eigenpairs of the augmented matrix $B \in \mathbb{R}^{(m+n) \times (m+n)}$ in Eqn.~\eqref{eq:B}. If $U^{\perp} \in \mathbb{R}^{m \times (m-n)}$ denotes a basis for the orthogonal complement subspace of $U\in\mathbb{R}^{m\times n}$; then,\\
\begin{equation}\label{eq:X}
X = \frac{1}{\sqrt{2}}
\begin{bmatrix}
V & -V & 0\\
U & U & \sqrt{2} U^{\perp} 
\end{bmatrix}\in\mathbb{R}^{(m+n)\times (m+n)},
\end{equation}
where $V\in\mathbb{R}^{n\times n}$, is an orthogonal matrix. With the initial vector guesses $\frac{1}{\sqrt{2}}[\tilde{\bs{v}}_i; \tilde{\bs{u}}_i]$, for $i=1, \dots, L$, of the eigensolver as follows from Eqn.~\eqref{eq:X}, we compute the ED of $B$ as
\begin{equation}\label{eqn:edB}
BX = X \ Diag(\lambda_1, ..., \lambda_n, -\lambda_1, ..., -\lambda_n, 0, ..., 0),
\end{equation}
where $0$ is repeated ($m-n$)-times in $Diag$. This approach can compute all eigenpairs accurately such that the residual norm is close to $\|A\|_2 \epsilon$. Let, $\left(\tilde{\lambda}^B_i, \tilde{\bs{x}}^B_i\in\mathbb{R}^{m+n}\right), i = 1, \dots, (m+n)$, be eigenpairs of $B$ approximated by applying the Rayleigh-Ritz scheme on the vector basis of $\left[\bs{x}^B_1, \dots, \bs{x}^B_{(m+n)}\right]$. Set, the singular triplets $\left(\tilde{\sigma}_i, \tilde{\bs{u}}_i, \tilde{\bs{v}}_i\right), i = 1, \dots, (m+n)$ such that $\tilde{\sigma}_i = \vert\tilde{\lambda}^B_i\vert$,  $\tilde{\bs{v}}_i = \tilde{\bs{x}}^B_i(1:n)$, and $\tilde{\bs{u}}_i = \tilde{\bs{x}}^B_i(n+1:m+n)$. PIME's convergence criterion is set to stop the execution when the residual norm of the singular value triplets of $B$ is less than $\sqrt{2}\|B\|_2\delta$, i.e.,
\begin{equation}\label{eq:convergeBC}
\|\tilde{\bs{r}}^B_i\|=\|B\tilde{\bs{x}}^B_i-\tilde{\lambda}^B_i\tilde{\bs{x}}_i\|_2\approx\sqrt{2}\|\tilde{\bs{r}}_i\|<\sqrt{2}\|B\|_2\delta,
\end{equation}
based on the assumption that $\|\tilde{\bs{x}}^B_i(1:n)\|_2\approx\|\tilde{\bs{x}}^B_i(n+1:m+n)\|_2$.

%%%%%%%%%%%%%%%%%%%%%%%%%
\begin{algorithm}[H]
\caption{Preconditioned Iterative Multimethod Eigensolver (PIME)\\
Input: data matrix ($A\in \mathbb{R}^{m\times n}$), required number of singular triplets ($L$), tolerance ($\delta$)\\
Output: approximated singular triplets $\{\tilde{\sigma_l}, \tilde{\mathbf{u}_l},\tilde{\mathbf{v}_l}, \}, l=1,...,L$.}
\label{alg:primme}
\begin{algorithmic}[1]
\item[] {\bf First stage:} working on $C$:
\State Set the convergence criterion according to Eqn.~\eqref{eq:convergeA}.
\State Compute eigenpairs $\left(\lambda^C_l, \bs{x}^C_l\right), l = 1, \dots, L$ using ED in Def.~\ref{def:ed} seeking $L$ largest eigenvalues of $C=A^TA$ (or $C = AA^T$)
\State Utilize the Rayleigh-Ritz scheme on the vector basis $\left[\bs{x}^C_1, \dots, \bs{x}^C_n\right]$ and approximate eigenpairs $\left(\tilde{\lambda}^C_l, \tilde{\bs{x}}^C_l\right), l = 1, \dots, L$
\State Set $\tilde{\sigma_l} = \vert \tilde{\lambda}_l^C \vert^{\frac{1}{2}}, \tilde{\mathbf{v}}_l = \tilde{\mathbf{x}}_l^C, \tilde{\mathbf{u}}_l = \mathcal{G} \tilde{\mathbf{v}}_l  \tilde{\sigma_l^{-1}}$
\If {all the singular triplets of $C$ converged, i.e., satisfy Eqn.~\eqref{eq:convergeC}}
	\State Return $\{\tilde{\sigma_l}, \tilde{\mathbf{u}_l}, \tilde{\mathbf{v}_l}, \}, l=1,...,L$
\EndIf
\vspace{5mm}
\item[] {\bf Second stage:} working on $B$:
\State Set the initial guesses as $\frac{1}{\sqrt{2}} 
\begin{bmatrix}
\tilde{\mathbf{u}_l}\\
\tilde{\mathbf{v}_l},
\end{bmatrix}
l=1,...,L
$
\State Set the convergence criterion according to Eqn.\eqref{eq:convergeBC}.  
\State Approximate the eigenpairs $\left(\tilde{\lambda}^B_l, \tilde{\bs{x}}^B_l\right), l = 1, \dots, L$ of $B$ given in Eqn.~\eqref{eq:B} by applying the Rayleigh-Ritz scheme on the vector basis of $X$ in Eqn.~\eqref{eq:X}.
\State Set $\tilde{\sigma}_l = \vert\tilde{\lambda}_l^B\vert$, $\tilde{\bs{v}}_l = \tilde{\bs{x}}^B_l(1:n)$, $\tilde{\bs{u}}_l = \tilde{\bs{x}}^B_l(n+1:m+n)$.
\State Return $\{\tilde{\sigma_l}, \tilde{\mathbf{u}_l}, \tilde{\mathbf{v}_l} \}, l=1,...,L$
\end{algorithmic}
\end{algorithm}
%%%%%%%%%%%%%%%%%%%%%%%%%

%%%%%%%%%%%%%%%%%%%%%%%%%
%%%%%%%          RSVD           %%%%%%%%%
\subsection{Randomized Singular Value Decomposition}\label{sec:rsvd}
Randomized Singular Value Decomposition (RSVD) \cite{halko2011finding}, combines probability theory with numerical linear algebra to develop an efficient, unbiased, and randomized algorithm to approximate SVD. RSVD efficiently approximates $L$ number of orthonormal vectors that span the range of  the given data matrix $A\in \mathbb{R}^{m\times n}$ so that such vectors enable an efficient approximation for SVD  of $A$. RSVD undergoes a two-stage computational process: 1) the approximation of orthonormal basis, denoted as $Q$, for the range of $A$; and 2) approximation of SVD of $A$ using such orthonormal basis $Q$. Randomness only occurs in Step 1 and Step 2 is deterministic for a given Q. The hybrid framework replacing the explicit singular vector computation by RSVD is denoted as GGD-RSVD.

\begin{itemize}
\item \textbf{First stage}
Here, the goal is to generate an orthonormal matrix $Q$ that consists of as few columns as possible such that
\begin{eqnarray}\label{eq:projector}
A \approx QQ^* A,
\end{eqnarray}
where $Q^*$ denotes the adjoint of $Q$. Draw $L$ Gaussian random columns $\bs{\omega}_1$, $\bs{\omega}_2$, \dots, $\bs{\omega}_L$ from $A_{m\times n}$. Then, project them sing the linear map $A$ such that $Y=A\Omega$ where $\Omega=[\bs{\omega}_1, \bs{\omega}_2, \dots, \bs{\omega}_L]$. Finally, find the orthonormal matrix $Q$ by using $QR$ factorization of $Y$, i.e., $Y=QR$. While having as few columns as possible in the basis matrix $Q$ increases efficiency, having more columns in it increases the accuracy of the approximation.

%%%%%%%%%%%
\begin{algorithm}[H]
\caption{
Randomized Singular Value Decomposition (RSVD) \cite{halko2011finding}.\\
Inputs: data matrix ($A\in \mathbb{R}^{m\times n}$) and the number of desired singular triplets ($L$).\\
Outputs: approximated singular triplets $\{\tilde{\sigma_l}, \tilde{\mathbf{u}_l},\tilde{\mathbf{v}_l}, \}, l=1,...,L$.}\label{alg:random}
\begin{algorithmic}[1]
\item[] {\bf First stage:}
\State Produce an $n \times L$-dimensional Gaussian matrix $\Omega$ by drawing columns of $A$ randomly.
\State Produce $Y = A \Omega$.
\State Generate an orthonormal matrix $Q$ using $QR$ factorization, i.e., $Y=QR$. 
\State Produce a matrix $Q$ such that its columns represent an orthonormal basis for the range of $Y$.
\vspace{5mm}
\item[]{\bf Second stage:}
\State Form $B = Q^* A$
\State Perform SVD of the $B = W \Sigma V^*$
\State Set $U = QW$
\State Return $\{\tilde{\sigma_i}, \tilde{\mathbf{u}_i}, \tilde{\mathbf{v}_i} \}, i=1,...,L$, where $\tilde{\sigma_i}$, $\tilde{\mathbf{u}_i}$, and $\tilde{\mathbf{v}_i}$ are the $i$-th diagonal entry of $\Sigma$, $i$-th column of $U$, and $i$-th column of $V^*$, respectively. 
\end{algorithmic}
\end{algorithm}
%%%%%%%%%%%

\item \textbf{Second stage}
Since the number of columns $L$ of the matrix $Q$ is significantly less than both the dimensions of $Q$, it is efficient to compute
\begin{eqnarray}\label{eq:B1}
B = Q^* A,
\end{eqnarray}
where $B$ consists of only $L$-many rows. Thus, it permits efficient SVD of $B$ as
\begin{eqnarray}\label{eq:W}
B = W \Sigma V^*
\end{eqnarray}
where the columns of both $W$ and $V$ are orthonormal, and $\Sigma_{L \times L}$ is a diagonal matrix whose entries are all non-negative. Let,  
\begin{eqnarray}\label{eq:U}
U = QW
\end{eqnarray}
and combining Eqns.~\eqref{eq:projector}--\eqref{eq:U} yields SVD of $A$ such that
\begin{eqnarray}\label{eq:comb}
A \approx QQ^* A = Q(B) = Q W \Sigma V^* =  U \Sigma V^*.
\end{eqnarray}
Note that, since $L$ is significantly less than both dimensions of $A$, $A$ has substantially more entries than any other matrix in RSVD.\\
\end{itemize}

%%%%%%%%%%%%%%%%%%%%%%%%%
%%% Performance Analysis   %%%%%%%%%
%%%%%%%%%%%%%%%%%%%%%%%%%
\section{Performance Analysis}\label{sec:analysis}
In the performance analysis, we analyze the proposed GGD-SVD and four of its variants, namely, GGD-MCLA, GGD-ALB, GGD-PIME, and GGD-RSVD, in terms of accuracy and computational time by implementing them on image denoising tasks. The scope of our experimental procedure is applying GGD-SVD and its four variants to three test images, namely Barbara, Cameraman, and Mandrill, that are distorted with three relative noise levels. The accuracy of the noise filtration is assessed by three performance metrics:  relative error (RE);  peak signal-to-noise ratio (PSNR); structural similarity index measure (SSIM). In the remainder of this section, we include three subsections on image preprocessing, performance metrics, and examples.

%%%%%%%%%%%%%%%%%%%%%%%%%
%%%%%        Gaussian Noise           %%%%%%%%
\subsection{Image preprocessing}\label{sec:gauss}
In the paper, we consider one of the most widely used Gaussian noise since such noise has practical implications as it is present in most images taken by present cameras.  Thus, we contaminate our test images with Gaussian noise and then attempt to filter them out using GGD frameworks presented in this paper. The \textit{Gaussing Probability Density Function} (GPDF) is assumed in the form: 
\begin{equation}\label{eqn:gauss}
P_g(z,\mu,\gamma) = {\frac{1}{\sigma \sqrt{2 \pi}} { e^{ \frac{ -{(z-\mu)}^2 } { 2 \sigma^2 } } } }.
\end{equation}
The symbols $z, \; \mu, \; \sigma$ in the above equation denote the gray-scale, mean of the distribution, and the standard deviation of the distribution, respectively. 

Suppose an $n\times n$ noise-free image is given and it is denoted as $\mathcal{I}_{n\times n}$. Consider a noise sample $\mathcal{N}_{n\times n}$ and from which we can find the GPDF given in Eqn.~\eqref{eqn:gauss} with $\mu=0$ and some $\sigma$. Then we can obtain a noisy image stemming from a simple additive rule as
\begin{equation}
\mathcal{U}_{n\times n}=\mathcal{I}_{n\times n}+\mathcal{N}_{n\times n}.
\end{equation}
Our analysis is based only on the gray images, hence we bring in a transformation of the colored images into gray images and such a process considers the average of the three color channels. The relative percentage noise, denoted as $\zeta$, can be computed by using 
\begin{equation}\label{eqn:noise}
\zeta=\frac{\|\mathcal{U}-\mathcal{I}\|_2}{\|\mathcal{I}\|_2} 100\%.
\end{equation}
The original noise-free image $\mathcal{I}_{100 \times 100}$ is enforced with different layers of noise intensities by using a variety of values of $\sigma$ of the GPDF in Eqn.~\eqref{eqn:gauss}  to construct different images with noise levels such as with $\zeta=30\%$, $40\%$, and $50\%$.  Then the entries in  $\mathcal{U}$ should be between 0-255 regardless of imposing the noise. Consequently, we adjust the rows of $\mathcal{U}$ by replacing anything below zero or above 255 with 255s.

%%%%%%%%%%%%%%%%%%%%%%%%%
%%%%%%%            Metrics            %%%%%%%%
\subsection{Performance metrics}\label{sec:metrics}
The performance of the GGD frameworks is assessed by three famous performance metrics RE, PSNR, and SSIM which we provide in Def.~\ref{def:re}, Def.~\ref{def:psnr}, and Def.~\ref{def:ssim}, respectively. While RE and PSNR measure two aspects of the quality of the image reconstruction, RE and PSNR are related since the term $\|\mathcal{I}-\tilde{\mathcal{U}}\|_2$ in RE is related to RMSE in PSNR by $\|\mathcal{I}-\tilde{\mathcal{U}}\|_2=n \times RMSE$. Thus, RE becomes a small value and PSNR becomes a large value in the case of $\mathcal{I}\approx\tilde{\mathcal{U}}$ that guarantees better denoising.

\begin{definition}\label{def:re}
Let, the two-dimensional matrix $\mathcal{I}$ represent an $n\times n$  reference image and $\tilde{\mathcal{U}}$ be any other image (from any area of interest).  The relative error (RE) of $\tilde{\mathcal{U}}$ in regard to  the image $\mathcal{I}$ is given as
\begin{equation}
RE(\mathcal{I},\tilde{\mathcal{U}}) = \frac{\|\mathcal{I}-\tilde{\mathcal{U}}\|_2}{\|\mathcal{I}\|_2}.
\end{equation}
From the above formula, it is clear that the values of RE vary between $0$ and $1$, this is likely due to the fact that the value of RE zero guarantees the pure similarity between $\mathcal{I}$ and $\hat{\mathcal{U}}$ while the value $1$ ensures perfect dissimilarity.
\end{definition}   

\begin{definition}\label{def:psnr}
Let, the two-dimensional matrix $\mathcal{I}_{n\times n}$ represent a reference image and $\tilde{\mathcal{U}}$ denote any other image. The \textit{Peak Signal to Noise Ratio} \cite{Hore2010} (PSNR) of the image $\tilde{\mathcal{U}}$ in reference to the image $\mathcal{I}$ is defined as  
\begin{equation}
PSNR(\mathcal{I},\tilde{\mathcal{U}}) = 20 \log_{10}\left(\frac{\max(\mathcal{I})}{RMSE(\mathcal{I},\tilde{\mathcal{U}})}\right).
\end{equation}
Here the symbol $\max(\mathcal{I})$ denotes the maximum pixel value of the image $\mathcal{I}$. Due to the fact the pixels in the images are represented in 8-bit digits, hence the $\max(\mathcal{I})$ is 255. It is clear the PSNR varies between zero and $\infty$, and the lower value zero established the pure dissimilarity between $\mathcal{I}$ and $\hat{\mathcal{U}}$ and the upper value $\infty$ establishes the perfect similarity.
\end{definition}   

\begin{definition}\label{def:ssim}
Let, the $n\times n$ and two-dimensional matrix $\mathcal{I}$ represent a reference image and $\tilde{\mathcal{U}}$  denotes an image from the field of interest. The \textit{Structural Similarity Index} (SSIM) \cite{Wang2004} is a measure of the image of the image $\tilde{\mathcal{U}}$ with respect to the reference image $\mathcal{I}$ is defined as
\begin{equation}
SSIM(\mathcal{I},\tilde{\mathcal{U}}) = I(\mathcal{I},\tilde{\mathcal{U}})C(\mathcal{I},\tilde{\mathcal{U}})S(\mathcal{I},\tilde{\mathcal{U}}),
\end{equation}
where
\begin{equation}
\begin{split}
I(\mathcal{I},\tilde{\mathcal{U}}) = \frac{2\mu_{\mathcal{I}}\mu_{\tilde{\mathcal{U}}}+c_1}{\mu^2_{\mathcal{I}}+\mu^2_{\tilde{\mathcal{U}}}+c_1},\\
C(\mathcal{I},\tilde{\mathcal{U}}) = \frac{2\sigma_{\mathcal{I}}\sigma_{\tilde{\mathcal{U}}}+c_2}{\sigma^2_{\mathcal{I}}+\sigma^2_{\tilde{\mathcal{U}}}+c_2},\\
S(\mathcal{I},\tilde{\mathcal{U}}) = \frac{\sigma_{\mathcal{I}\tilde{\mathcal{U}}}+c_3}{\sigma_{\mathcal{I}}\sigma_{\tilde{\mathcal{U}}}+c_3}.\\
\end{split}
\end{equation}
Here, $\mu_{\mathcal{I}}$ and $\mu_{\tilde{\mathcal{U}}}$ are the means of $\mathcal{I}$ and $\tilde{\mathcal{U}}$, respectively; $\sigma_{\mathcal{I}}$ and $\sigma_{\tilde{\mathcal{U}}}$ are standard deviations of $\mathcal{I}$ and $\tilde{\mathcal{U}}$, respectively; and  $\sigma_{\mathcal{I}\tilde{\mathcal{U}}}$ is the covariance between $\mathcal{I}$ and $\tilde{\mathcal{U}}$. Moreover, $ 0< c_1, \;  c_2, \; c_3 \ll 1$ are like regularization parameters to avoid division by zero. From the above formula, it is clear that the values of SSIM lie within the interval $(-1, \; 1)$ so that the extreme values indicate pure dissimilarity and perfect similarity. 
\end{definition}      

%%%%%%%%%%%%%%%%%%%%%%%%%
%%%%%%%            Examples            %%%%%%%%
\subsection{Examples}
\begin{figure}[!tp] 
\includegraphics[width = 1\textwidth]{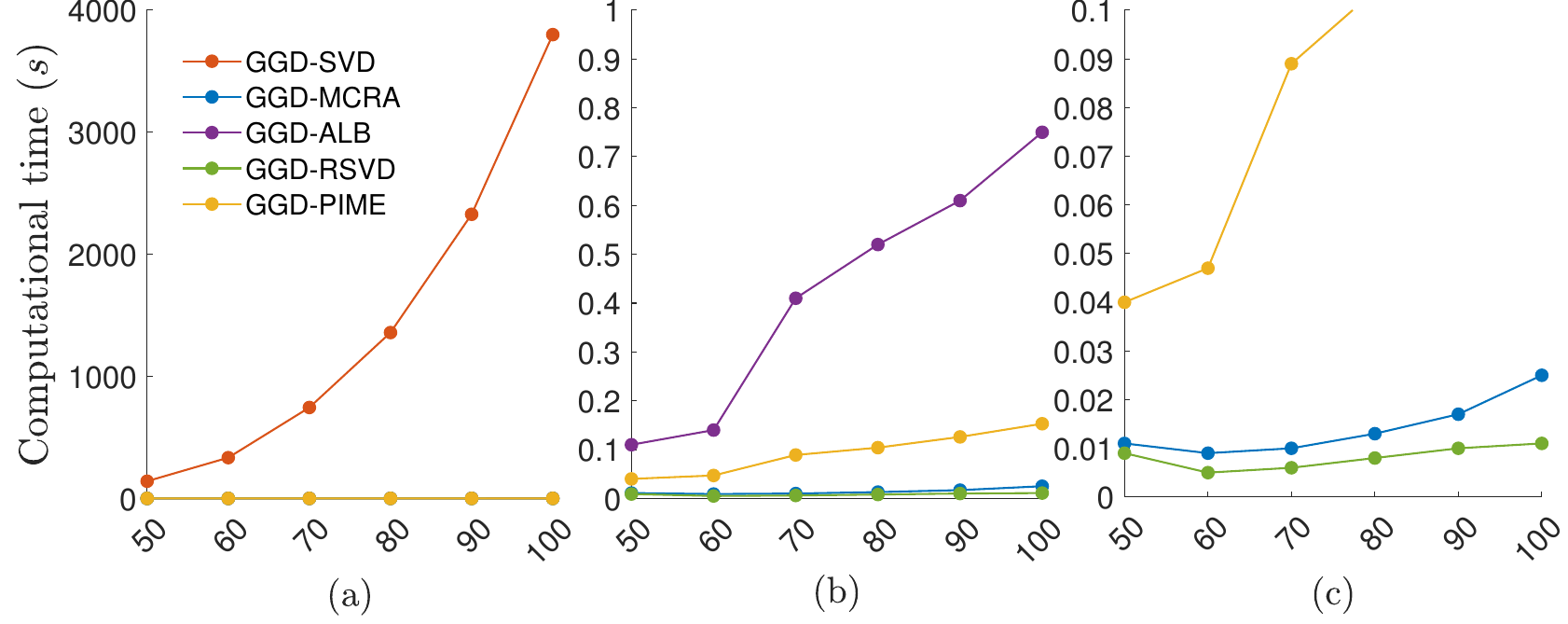}\vspace{-5mm}
\caption{Computational time, in seconds ($s$), of GGD-SVD and four of its hybrid variants with respect to the length of one side of a square-shaped image, i.e, $n$ for an image of $n\times n$. First, we create six versions of the image Barbara with the sizes $n=50, 60, 70, 80, 90$, and $100$, and impose an additive Gaussian noise sample with relative percentage noise of 30\%. We executed all five frameworks over these test images on a computer with the configuration of $11^{th}$ Generation Intel Core i7-1165G7, 4 cores each with 4.70 GHz, 8 GB DDR4 3200MHz RAM. (a) Computational time attained by each framework of denoising six test images where the plots of GGD-MCRA, GGD-ALB, GGD-RSVD, and GGD-PIME are overlapped. While (b) shows the first cropped version of (a) which limits the vertical axis at 1$s$, (c) shows the second cropped version of (a) which limits the vertical axis at 0.1$s$.} 
\label{fig:time}
\end{figure}

In this subsection, we briefly describe the implementation of the proposed baseline algorithm, GGD-SVD, and its four variants to denoise the noisy versions of three benchmark computer vision test images, Barbara, cameraman, and mandrill available in Ref.~\cite{Moghaddam2014}. In order to assess the efficacy of the proposed framework, we implemented five frameworks on images of variable sizes, and in all our calculations the computer architecture was kept the same.  Here, we work on square-shaped images of size $n\times n$ for simplicity. First, we create six versions of the image Barbara with the lengths $n=50, 60, 70, 80, 90, 100$ and impose an additive Gaussian noise sample with a relative percentage noise of 30\%. Then, we execute GGD-SVD and its four hybrid frameworks over the images of $n=50, 60, 70, 80, 90$, and $100$ with some arbitrary parameter combination of $(\delta, \rho, L) =(10, 5, 15)$, and compute the computational time for denoising. Fig. \ref{fig:time} showing the computational time, in seconds ($s$), with respect to image length ($n$) justifies that the hybrid versions of GGD that are made by incorporating SVD approximation techniques into GGD attain significantly less computational time. We observe that the increasing order of the computational costs of the four variants are GGD-RSVD, GGD-MCRA, GGD-PIME, and GGD-ALB.

\begin{table}[!htp] 
\caption{PSNR and RE for the denoising of GGD-SVD and its four hybrid frameworks, GGD-MCLA, GGD-ALB, GGD-PIME, and GGD-RSVD. Three test images, namely, Barbara, cameraman, and mandrill, are imposed with three Gaussian relative noise levels, $\zeta=$ 20\%, 30\%, and 40\%.  Each noisy image is denoised using each method along with its best parameter combination $(\delta, \rho, L)$ that is optimized with respect to PSNR. The best parameter combinations are shown in parenthesis, the REs for those combinations are shown in square brackets, and the PSNRs for those combinations are shown in other values. For each noisy image, we used the coloring red, blue, and green to indicate the method with the best, second best, and third best performance. The parameter space for the optimization is $\delta=8, 10, 12$, $\rho=3, 5, 7$, and $L=15, 20, 25$ for $\zeta=$20\%; $\delta=10, 12, 14$, $\rho=5, 7, 9$, and $L=15, 20, 25$ for $\zeta=$30\%; and $\delta=12, 14, 16$, $\rho=7, 9, 11$, and $L=15, 20, 25$ for $\zeta=$40\%.}
\label{tab:psnr}
\centering
 \begin{tabular}{|c|c|c c c c c|} 
 \hline
 Data                                   &   $\zeta (\%)$   & GGD-SVD & GGD-MCLA &       GGD-ALB & GGD-PIME & GGD-RSVD\\
 \hline\hline
\multirow{9}{*}{\rotatebox[origin=c]{90}{Barbara}} & \multirow{3}{*}{20} & {\color{green}22.84}  & 22.22  &  {\color{blue}22.88 } &  {\color{blue}22.88 } &  {\color{red}22.89 }\\ 
				      &				& [0.1358]  	& [0.1459] 	&  [0.1353] 	&  [0.1351] 	&  [0.1351]\\
                                      &                                 & $(10,5,20)$      & $(10,7,15)$     & $(10,7,15)$      & $(10,7,15)$       & $(8,7,15)$ \\ 
\cline{2-7}
                                      & \multirow{3}{*}{30} & {\color{red}22.32 } & 21.39  & {\color{green}22.09 } & {\color{green}22.09 } & {\color{blue}22.17}\\ 
                                      &				& [0.1443] 	& [0.1605] 	& [0.1481] 	& [0.1481] 	& [0.1468]\\ 
                                      &                                 & $(10,5,15)$      & $(10,7,20)$     & $(10,7,15)$      & $(10,7,15)$       & $(12,9,15)$ \\ 
\cline{2-7}
                                      & \multirow{3}{*}{40} &  {\color{green}20.80 }  & 20.27  & {\color{blue}21.02 } & {\color{blue}21.02 } &  {\color{red}21.26 }\\ 
                                      &				&  [0.1719]  	& [0.1826] 	& [0.1676] 	& [0.1676] 	&  [0.1629]\\ 
                                      &                                 & $(12,7,15)$      & $(14,7,25)$     & $(16,9,15)$      & $(16,9,15)$       & $(16,9,15)$ \\ 
 \hline\hline
\multirow{9}{*}{\rotatebox[origin=c]{90}{Cameraman}} & \multirow{3}{*}{20}  &  20.94             &  {\color{red}22.91} &  {\color{green}21.07 } & {\color{blue}22.28 } & {\color{blue}22.28 }\\ 
					&				&  [0.1721]              		&  [0.1371] 	&  [0.1695] 	& [0.1474] 	& [0.1474]\\ 
                                      &                                 &    $(10,3,25)$                 & $(8,3,15)$     & $(12,7,15)$      & $(10,3,15)$       & $(10,3,15)$ \\ 
\cline{2-7}
                                      & \multirow{3}{*}{30} & 19.61   &  {\color{green}20.78 } & 20.54  &  {\color{red}21.20 } & {\color{blue}20.97 }\\ 
                                      &				&	[0.2005]  	&  [0.1753] 	& [0.1801] 	&  [0.1670] 	& [0.1714]\\ 
                                      &                                 & $(12,7,15)$      & $(14,5,25)$     & $(14,7,15)$      & $(14,5,15)$       & $(12,5,15)$ \\ 
\cline{2-7}
                                      & \multirow{3}{*}{40} & 17.91   &  {\color{green}19.29} & {\color{blue}19.60 } & {\color{blue}19.60 } &  {\color{red}19.78}\\ 
                                      &				&	[0.12439]  	&  [0.2081] 	& [0.2007] 	&[0.2007] 		&  [0.1968]\\ 
                                      &                                 & $(16,11,25)$      & $(14,7,15)$     & $(16,7,15)$      & $(16,7,15)$       & $(16,9,20)$ \\ 
 \hline\hline
\multirow{9}{*}{\rotatebox[origin=c]{90}{Mandrill}} & \multirow{3}{*}{20}  &  {\color{green}21.07 }    & {\color{blue}21.36 } & 20.29  &  {\color{red}21.38 } & 21.15 \\ 
					&				&	[0.1530]   	& [0.1481] 	& [0.1675] 	&  [0.1477] 	&  [0.1517]\\ 
                                      &                                 & $(8,3,25)$         & $(8,7,15)$     & $(12,7,15)$      & $(8,3,15)$       & $(8,3,15)$ \\ 
\cline{2-7}
                                      & \multirow{3}{*}{30} & {\color{blue}19.02}  & 18.80&  {\color{green}18.98 } &  {\color{green}18.98 } &  {\color{red}19.03 }\\ 
                                      &				& [0.1939]  	&  [0.1989] 	&  [0.1947] 	&  [0.1947] 	&  [0.1937]\\ 
                                      &                                 & $(12,7,20)$      & $(10,7,25)$     & $(10,7,15)$      & $(10,7,15)$       & $(14,7,15)$ \\ 
\cline{2-7}
                                      & \multirow{3}{*}{40} & 14.98  &  {\color{green}18.93 } & {\color{blue}19.20 } & {\color{blue}19.20 } &  {\color{red}19.23 }\\ 
                                      &				&	[0.1948]  	&  [0.1959] 	& [0.1899] 	& [0.1899] 	&  [0.1891]\\ 
                                      &                                 & $(14,7,25)$      & $(14,9,20)$     & $(16,9,15)$      & $(16,9,15)$       & $(14,7,15)$ \\ 
 \hline
 \end{tabular}
\end{table}

\begin{figure}[!htp]
\includegraphics[width = 1\linewidth]{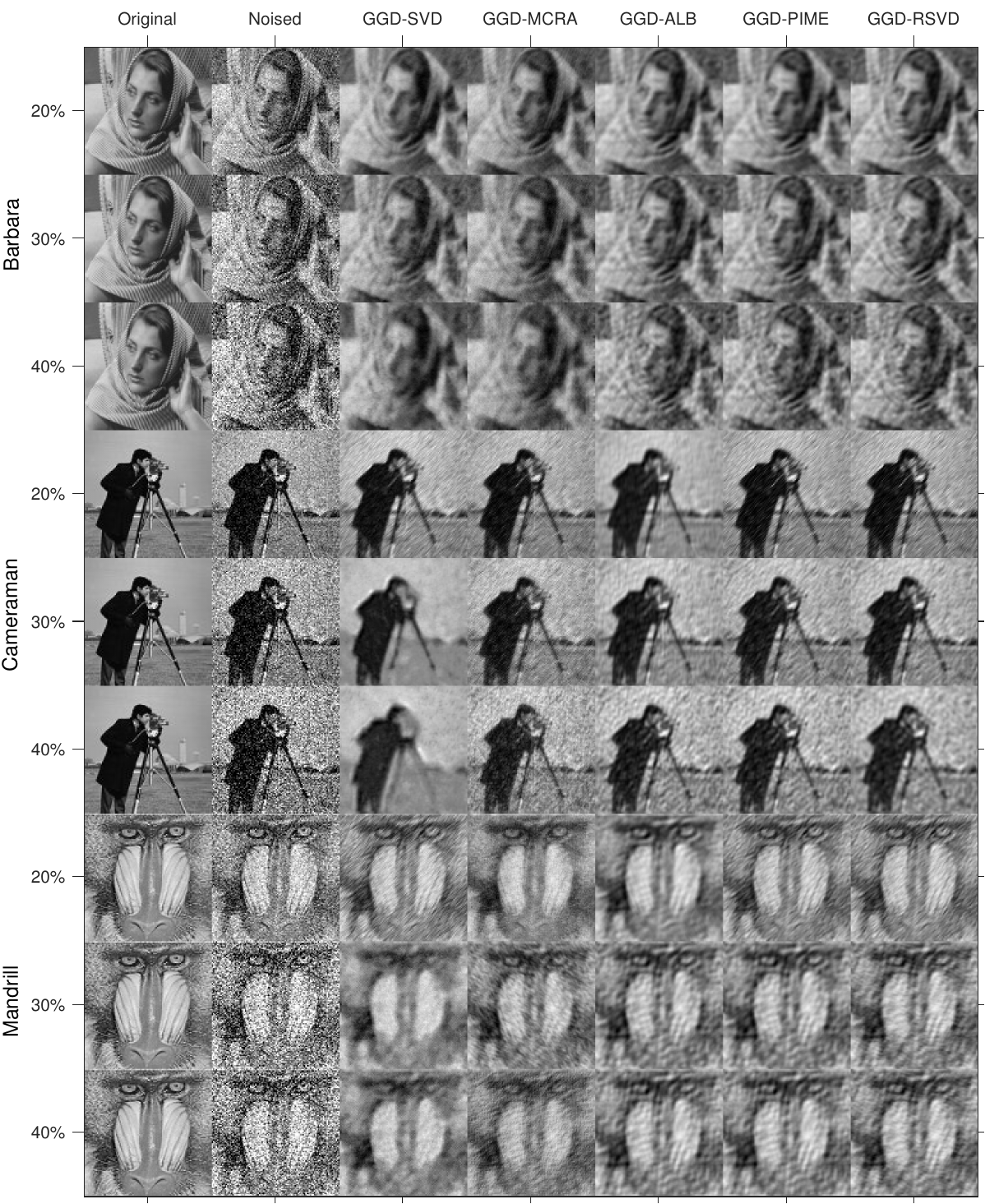}\vspace{-5mm}
\caption{Denoising images using GGD-SVD and four of its hybrid frameworks, GGD-MCLA, GGD-ALB, GGD-PIME, and GGD-RSVD. Three test images, namely, Barbara, cameraman, and mandrill, are imposed with three Gaussian relative noise levels, $\zeta=$ 20\%, 30\%, and 40\%.  Each noisy image is denoised using each method along with its best parameter combination $(\delta, \rho, L)$, shown in Table~\ref{tab:psnr}, which is optimized with respect to PSNR.} 
\label{fig:psnr}
\end{figure}

\begin{figure}[htp]
\includegraphics[width = 1\linewidth]{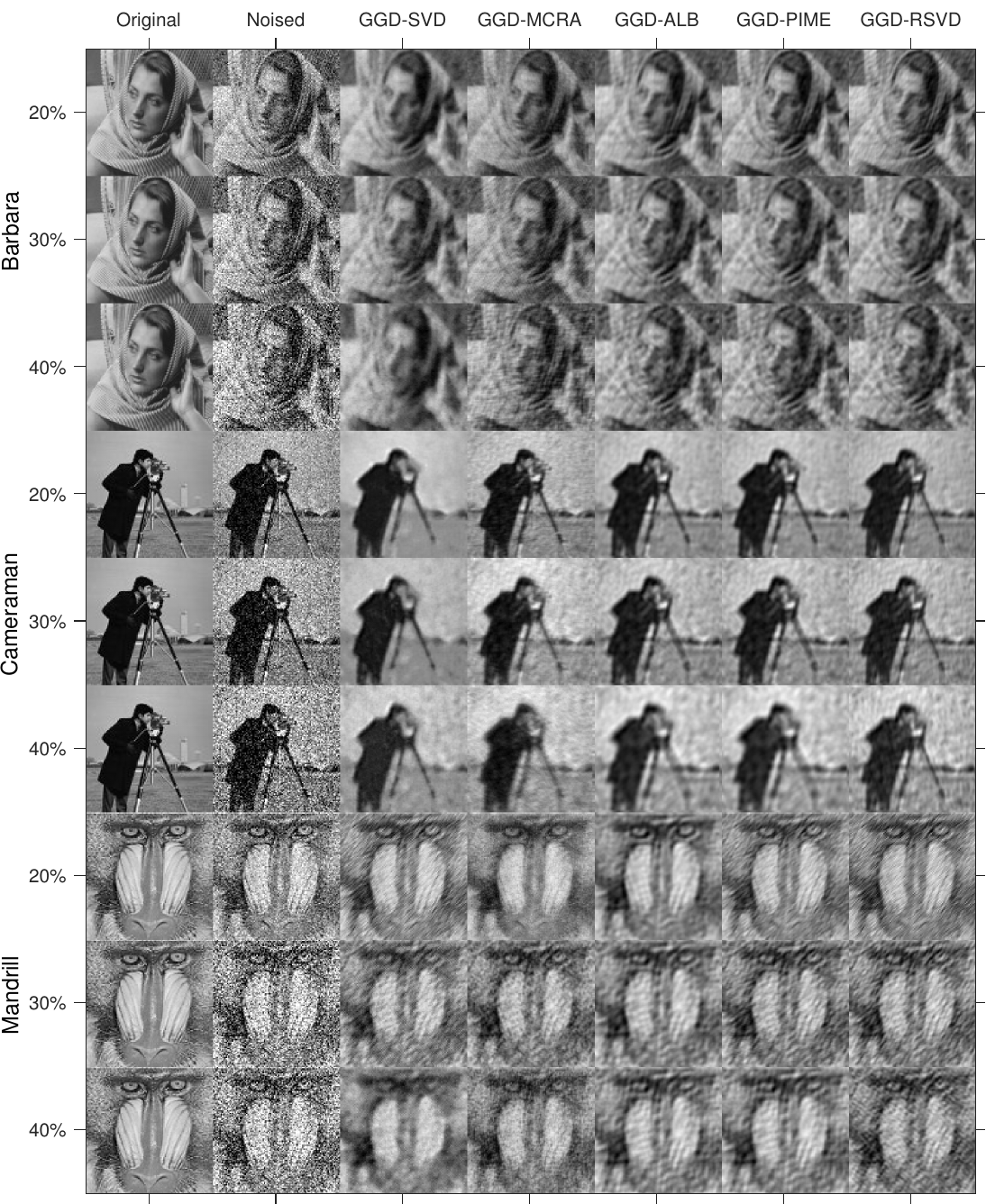}\vspace{-5mm}
\caption{Denoising images using GGD-SVD and four of its hybrid frameworks, GGD-MCLA, GGD-ALB, GGD-PIME, and GGD-RSVD,       with optimized parameters with respect to SSIM. Three test images, namely, Barbara, cameraman, and mandrill, are imposed with three Gaussian relative noise levels, $\zeta=$ 20\%, 30\%, and 40\%.  Each noisy image is denoised using each method along with its best parameter combination $(\delta, \rho, L)$, shown in Table~\ref{tab:ssim}, which is optimized with respect to SSIM.} 
\label{fig:ssim}
\end{figure}

Now, we asses the denoising performance of GGD-SVD and four of its hybrid frameworks by implementing them on three test images Barbara, cameraman, and mandrill. Our parameter space is $\delta=8, 10, 12$, $\rho=3, 5, 7$, and $L=15, 20, 25$ for $\zeta=$20\%; $\delta=10, 12, 14$, $\rho=5, 7, 9$, and $L=15, 20, 25$ for $\zeta=$30\%; and $\delta=12, 14, 16$, $\rho=7, 9, 11$, and $L=15, 20, 25$ for $\zeta=$40\%. First, we impose three noise levels, $\zeta=$ 20\%, 30\%, and 40\% into each of the three test images. Then, for each test image and the noise level, we implement GGD-SVD and the four of its hybrid frameworks with all the parameter combinations, i.e., $\delta$'s, $\rho$'s, and $\zeta$'s. Using three performance metrics (RE, PSNR, and SSIM), we calculate the similarity between the noise-free original test image and its denoised version. In all the computations, the best-denoised version of the test image and noise level is chosen in reference to both PSNR and SSIM.Table~\ref{tab:psnr} presents both RE and PSNR along with optimized parameter values with respect to PSNR, as they are related so that the best denoising is predicated by small RE and big PSNR. Fig.~\ref{fig:psnr} shows the images denoised by each method with their optimized parameter values. We observe that RSVD attains the best denoising performance with respect to PSNR while PIME attains the second best. The performance of  GGD-SVD, GGD-MCLA, and GGD-ALB are mostly similar. The performance, assessed using PSNR, of all five methods decreases as the noise contamination increases. Table~\ref{tab:ssim} presents SSIM and optimized parameter values with respect to SSIM. Fig.~\ref{fig:ssim} shows the images denoised by each method with their optimized parameter values. We observe in Table~\ref{tab:ssim} that while PIME attains the best denoising performance with respect to SSIM, all the other frameworks' performances are mostly similar. The structural similarity between the true and the denoised images decreases for all five methods as the noise contamination increases.

\begin{table}[htp]
\caption{SSIM for the denoising of GGD-SVD and its four hybrid frameworks, GGD-MCLA, GGD-ALB, GGD-PIME, and GD-RSVD. Three test images, namely, Barbara, cameraman, and mandrill, are imposed with three Gaussian relative noise levels, $\zeta=$ 20\%, 30\%, and 40\%.  Each noisy image is denoised using each method along with its best parameter combination $(\delta, \rho, L)$ that is optimized with respect to SSIM. The best parameter combinations are shown in parenthesis and the SSIMs for those combinations are shown in other values. For each noisy image, we used the coloring red, blue, and green to indicate the method with the best, second best, and third best performance. The parameter space for the optimization is $\delta=8, 10, 12$, $\rho=3, 5, 7$, and $L=15, 20, 25$ for $\zeta=$20\%; $\delta=10, 12, 14$, $\rho=5, 7, 9$, and $L=15, 20, 25$ for $\zeta=$30\%; and $\delta=12, 14, 16$, $\rho=7, 9, 11$, and $L=15, 20, 25$ for $\zeta=$40\%.}
\label{tab:ssim}
\centering
 \begin{tabular}{|c|c| c c c c c|} 
 \hline
 Data                                   & $\zeta (\%)$     & GGD-SVD & GGD-MCLA &       GGD-ALB & GGD-PIME & GGD-RSVD\\
 \hline\hline
\multirow{6}{*}{\rotatebox[origin=c]{90}{Barbara}} & \multirow{2}{*}{20} & 0.6395  & {\color{green}0.6527} & 0.6398 & {\color{red}0.6638} &  {\color{blue}0.6536} \\ 
                                      &                                 & $(12,5,15)$      & $(8,5,25)$     & $(12,7,15)$      & $(8,5,15)$       & $(12,5,15)$ \\ 
\cline{2-7}
                                      & \multirow{2}{*}{30} & 0.5848  & {\color{blue}0.5864}  & {\color{green}0.5858} & {\color{green}0.5858} &  {\color{red}0.5882} \\ 
                                      &                                 & $(10,5,15)$      & $(10,5,25)$     & $(14,9,15)$      & $(14,9,15)$       & $(14,9,15)$ \\ 
\cline{2-7}
                                      & \multirow{2}{*}{40} & 0.5069  & {\color{green}0.5168} & {\color{red}0.5310} &  {\color{red}0.5310} & {\color{blue}0.5309} \\ 
                                      &                                 & $(12,7,15)$      & $(16,7,20)$     & $(16,9,15)$      & $(16,9,15)$       & $(14,9,15)$ \\ 
 \hline\hline
\multirow{6}{*}{\rotatebox[origin=c]{90}{Cameraman}} & \multirow{2}{*}{20}  &  {\color{red}0.6359}     &0.5991  & {\color{blue}0.6256} & {\color{blue}0.6256} & {\color{green}0.6221} \\ 
                                      &                                 &     $(8,7,25)$                & $(8,5,25)$     & $(8,7,15)$      & $(8,7,15)$       & $(12,7,15)$ \\ 
\cline{2-7}
                                      & \multirow{2}{*}{30} &   {\color{red}0.5784} & 0.5063 & {\color{green}0.5170}  & {\color{green}0.5170}  & {\color{blue}0.5207} \\ 
                                      &                                 & $(12,7,15)$      & $(10,7,20)$     & $(10,7,15)$      & $(10,7,15)$       & $(12,7,15)$ \\ 
\cline{2-7}
                                      & \multirow{2}{*}{40} &   {\color{red}0.4820} & {\color{green}0.4542}  & {\color{blue}0.4550}  & {\color{blue}0.4550}  &0.4485 \\ 
                                      &                                 & $(12,7,15)$      & $(16,11,20)$     & $(12,11,15)$      & $(12,11,15)$       & $(12,9,15)]$ \\ 
 \hline\hline
\multirow{6}{*}{\rotatebox[origin=c]{90}{Mandrill}} & \multirow{2}{*}{20}  &   {\color{green}0.5847} &  {\color{red}0.6326} &0.5239  & {\color{blue}0.6050}  &0.5825 \\ 
                                      &                                 & $(10,3,15)$         & $(8,7,15)$     & $(8,7,15)$      & $(8,3,15)$       & $(10,3,15)$ \\ 
\cline{2-7}
                                      & \multirow{2}{*}{30} &  0.4144&  {\color{red}0.4579} &0.4250  & {\color{blue}0.4435}  & {\color{green}0.4427} \\ 
                                      &                                 & $(14,5,20)$      & $(12,5,20)$     & $(10,7,15)$      & $(12,5,15)$       & $(12,5,15)$ \\ 
\cline{2-7}
                                      & \multirow{2}{*}{40} &  0.3643 &  {\color{red}0.4536}  & {\color{green}0.4173}  & {\color{green}0.4173} & {\color{blue}0.4503}\\ 
                                      &                                 & $(16,7,20)$      & $(12,7,15)$     & $(16,9,15)$      & $(16,9,15)$       & $(16,7,15)$ \\ 
 \hline
 \end{tabular}
\end{table}

%%%%%%%%%%%%%%%%%%%%%%%%%
%%%%%%%      Conclusion         %%%%%%%%%
%%%%%%%%%%%%%%%%%%%%%%%%%
\section{Conclusion}\label{sec:conclusion}
Filtering out noise from an image while preserving image features is an arduous task as noise blends into features. While the algebraic image processing frameworks are sometimes inefficient for this denoising task as they may require processing of matrices of order equivalent to some power of the order of the original image, the neural network image processing frameworks are sometimes not robust as they require a lot of similar training samples. Thus, here we presented a robust image denoising framework GGD-SVD, and four of its efficient variants, GGD-MCLA, GGD-ALB, GGD-PIME, and GGD-RSVD, made by replacing explicit SVD computation by singular vectors approximations. The performance analysis was based on three test images each imposed with three diverse nose levels. In the current communication, we have done a visual comparison and also by three similarity metrics, namely, RE, PSNR, and SSIM. 

The conventional SVD encounters a computational complexity of $\mathcal{O}(mn \min(m, n))$ for a matrix of order $m \times n$. Denoising of the image requires GGD-SVD to compute some, say $L$, prominent singular values of patches' geodesic distance matrix of size $n^2\times n^2$ using SVD that encounters computational complexity of $\mathcal{O}(n^6)$, which makes the complexity of GGD-SVD to be $\mathcal{O}(n^6)$. The approximation methods that we incorporated were  Monte Carlo Low-Rank Approximation (MCLA), Augmented Lanczos Bidiagonalization (ALB),  Preconditioned Iterative Multimethod Eigensolver (PIME), and Randomized Singular Value Decomposition (RSVD). While the conventional approximation techniques, MCLA, ALB, and PIME encounter computational complexity of $\mathcal{O}(mnL)$, the randomization-based approximation method RSDV encounters only $\mathcal{O}(mn\log(L))$. Compared to GGD-SVD, all of its hybrid methods significantly reduced the computational time and the order of the computational time from the best hybrid framework to the worst is GGD-RSVD, GGD-MCRA, GGD-PIME, and GGD-ALB. Moreover, we observed that the computational time differences of GGD-RSVD and GGD-MCRA are not very significant. While the PSNR values provide that GGD-RSVD performs the best and GGD-PIME performs the next best, SSIM values provide that all four hybrid frameworks perform similarly. If we consider both the computational efficiency and denoising quality, while GGD-RSVD performs the best, GGD-MCRA performs next. 

We have been developing deep learning techniques for data imputation in computer vision \cite{gajamannage2021reconstruction} and time series forecasting in financial markets \cite{gajamannage2022real}. We are planning to extend our data imputation and time series forecasting techniques along with this efficient image denoising technique so that the extended framework is capable of operating on partially observed noisy image sequences. Such a framework is capable of denoising a video that contains some objects that are partially observed because of processes such as occlusion or change of appearance. The applications of such techniques could include learning the dynamics of animal groups in the wild, analyzing pedestrians on streets, counting vehicles on roads, and tracking sports players on courts \cite{Vicsek12}. Utilization of a single hybrid framework is favorable rather than three separate frameworks each for denoising, tracking, and trajectory imputation, as this hybrid framework increases both the computational efficiency and robustness. Future work also includes comparing these hybrid GGD frameworks with state-of-the-art deep learning image restoration tools such as Generative Adversarial Network \cite{goodfellow2014generative}, U-Net \cite{ronneberger2015u}. However, the use of deep learning techniques to learn image features requires ground truth, and it could be a limitation of such a denoising method as ground truth is not always available. The hybrid GGD frameworks presented in this paper are only capable of processing gray-color images; thus, we are planning to extend these four methods in the future such that they are capable of processing color images. For that, first, we decompose the given noisy color image into its three color channels. Then, each channel is denoised with our hybrid GGD separately with its best parameter values. Finally, we merge the denoised color channels and produce the denoised color image.

In this paper, we presented a robust patch-based image denoising technique, GGD-SVD, that uses eigenvectors of the geodesics' Gramian matrix computed using SVD. The computational cost of GGD-SVD encountered at its SVD step is eliminated to increase efficnecy by replacing explicit singular vector computation with four singular vector approximation frameworks, namely, MCLA, ALB, PIME, and RSVD. Both, the computational time comparison and denoising performance comparison evidence that the hybrid GGD framework GGD-RSVD is the most efficient and robust scheme among all the other techniques. 

%%%%%%%%%%%%%%%%%%%%%%%
\bibliographystyle{elsarticle-num} 
%\bibliography{ref/refAproxSVD}

\end{document}